\documentclass{aa}  

\usepackage{graphicx}
\usepackage{txfonts}
\usepackage{hyperref}
\hypersetup{
    colorlinks=true,
    linkcolor=blue,
    citecolor=blue
    }
\defcitealias{bb:10}{BB10}
\defcitealias{2023arXiv230611100D}{D24}
\defcitealias{1999AA...351..459C}{C99}
\defcitealias{2005AA...430...83C}{C05}
\defcitealias{2008AA...479...49B}{B08}
\defcitealias{2009AA...499..653B}{B09}
\defcitealias{2011MNRAS.412.1441L}{L11}
\defcitealias{2012ApJ...757...70D}{D12}
\defcitealias{2014ApJ...792..135T}{T14}
\defcitealias{2015MNRAS.450..905G}{G15}
\defcitealias{2015AA...584A..62C}{C15}
\defcitealias{2021MNRAS.500.5142F}{F21}
\defcitealias{2015MNRAS.450..905G}{G15}
\defcitealias{2020ApJ...904...35P}{P20}
\defcitealias{2019MNRAS.489.5802V}{V19}
\defcitealias{2012Natur.491..228C}{C12}
\defcitealias{2013MNRAS.431..912Q}{Q13}
\defcitealias{2017MNRAS.464.3568P}{P17}
\defcitealias{2019ApJS..241...16M}{M19}

\begin{document} 

   \title{Supernova rates and luminosity functions from ASAS-SN II: 2014–2017 core-collapse supernovae and their subtypes}

   % \subtitle{}

   \author{T. Pessi
          \inst{1,2}
          \and
          D. D. Desai
          \inst{3}
          \and
          J. L. Prieto
          \inst{2,4}
          \and 
          C. S. Kochanek
          \inst{5,6}
          \and 
          B. J. Shappee
          \inst{3}
          \and 
          J. P. Anderson
          \inst{1,4}
          \and 
          J.~F.~Beacom
          \inst{5,6,7}
          \and 
          Subo Dong 
          \inst{8,9,10}
          \and 
          K. Z. Stanek
          \inst{5,6}
          \and 
          T. A. Thompson
          \inst{5,6,7}
          }

   \institute{European Southern Observatory, Alonso de Córdova 3107, Vitacura, Casilla 19001, Santiago, Chile\\
   \email{Thallis.Pessi@eso.org}
   \and 
   Instituto de Estudios Astrof\'isicos, Facultad de Ingenier\'ia y Ciencias, Universidad Diego Portales, Av. Ej\'ercito Libertador 441, Santiago, Chile
   \and
   Institute for Astronomy, University of Hawai‘i at Mānoa, 2680 Woodlawn Drive, Honolulu, HI 96822, USA
   \and 
   Millennium Institute of Astrophysics MAS, Nuncio Monsenor Sotero Sanz 100, Off. 104, Providencia, Santiago, Chile
   \and 
    Department of Astronomy, The Ohio State University, 140 West 18th Avenue, Columbus, OH 43210, USA
    \and 
    Center for Cosmology and AstroParticle Physics, The Ohio State University, 191 West Woodruff Avenue, Columbus, OH 43210, USA
    \and 
    Department of Physics, The Ohio State University, 191 W. Woodruff Ave., Columbus, OH 43210, USA
    \and 
    Department of Astronomy, School of Physics, Peking University, Yiheyuan Rd. 5, Haidian District, Beijing 100871, China
    \and
    Kavli Institute for Astronomy and Astrophysics, Peking University, Yi He Yuan Road 5, Hai Dian District, Beijing 100871, PR China
    \and 
    National Astronomical Observatories, Chinese Academy of Science, 20A Datun Road, Chaoyang District, Beijing 100101, China
             }

   % \date{Received September 15, 1996; accepted March 16, 1997}

% \abstract{}{}{}{}{}
% 5 {} token are mandatory

\abstract
   {}
   {The volumetric rates and luminosity functions (LFs) of core-collapse supernovae (ccSN) and their subtypes are important for understanding the cosmic history of star formation and the buildup of ccSNe products. 
   To estimate these rates, we used data of nearby ccSNe discovered by the All-Sky Automated Survey for Supernovae (ASAS-SN) from 2014 to 2017, when all observations were made in the $V$ band.
   % We present volumetric rates and LFs of nearby core-collapse supernovae (ccSNe) and their subtypes from the All-Sky Automated Survey for Supernovae (ASAS-SN).
   }
   {The sample is composed of 174 discovered or recovered events, with high spectroscopic completeness from follow-up observations. 
   This allowed us to obtain a statistically precise and systematically robust estimate of nearby rates for ccSNe and their subtypes. The volumetric rates were estimated by correcting the observed number of events for survey completeness, which was estimated through injection recovery simulations using ccSN light curves.}
   {We find a total volumetric rate for ccSNe of $7.0^{+1.0}_{-0.9} \times 10^{-5} \ \textrm{yr}^{-1} \ \textrm{Mpc}^{-3} \ h^{3}_{70}$, at a median redshift of 0.0149, for absolute magnitudes at peak $M_{V,peak} \leq -14$ mag. This result is in agreement with previous local volumetric rates.
   We obtain volumetric rates for the different ccSN subtypes (II, IIn, IIb, Ib, Ic, Ibn, and Ic-BL), and find that the relative fractions of Type II, stripped-envelope, and interacting ccSNe are $63.2\%$, $32.3\%$, and $4.4\%$, respectively. 
   We also estimate a volumetric rate for superluminous SNe of $1.5^{+4.4}_{-1.1} \ \textrm{yr}^{-1} \ \textrm{Gpc}^{-3} \ h^{3}_{70}$, corresponding to a fraction of $0.002\%$ of the total ccSN rate. 
   We produced intrinsic $V$-band LFs of ccSNe and their subtypes, and show that ccSN rates steadily decline for increasing luminosities.
   % and find that ccSNe occur with volumetric rates of $\sim 0.9-2.4 \times 10^{-5} \ \textrm{yr}^{-1} \ \textrm{Mpc}^{-3} \ h^{3}_{70} \ \textrm{mag}^{-1} $, for a $M_{V,peak}$ range of $-18.0$ to $-15.5$ mag, respectively, which falls to $\sim 1.8\times 10^{-8} \ \textrm{yr}^{-1} \ \textrm{Mpc}^{-3} \ h^{3}_{70} \ \textrm{mag}^{-1} $ at $M_{V,peak} \approx -21.1$ mag.
   We further investigated the specific ccSN rate as a function of their host galaxy stellar mass and find that the rate decreases with increasing stellar mass, with significantly higher rates at lower mass galaxies (log $M_* < 9.0$ M$_\odot$).}
   {}


   \maketitle
%
%-------------------------------------------------------------------

\section{Introduction}

Core-collapse supernovae (ccSNe) are the end-of-life explosions of massive ($\geq 8$ M$_\odot$) stars \citep{1979NuPhA.324..487B, 1986ARA&A..24..205W, 1989ARA&A..27..629A, 2009ARA&A..47...63S}. 
They are responsible for dramatic changes in the evolution of galaxies, triggering or inhibiting the formation of new stars \citep{1986A&A...154..279M, 2010ApJ...714L.275M}. 
They are also one of the main formation channels of dust \citep{2009ASPC..414...43K} and heavy elements, driving the chemical enrichment of galaxies \citep[e.g.,][]{2011ApJ...729...16K, 2013RvMP...85..809K}. 
Due to their short delay time, they directly trace the star-formation (SF) rate \citep[e.g.,][]{2019MNRAS.488.5300C} and can be used as a tracer to study how SF changes with the physical properties of stellar populations, such as age and metallicity. 

Core-collapse supernovae can be broadly divided between hydrogen-rich Type II and hydrogen-poor, or stripped-envelope (SE) SNe. 
The SNe Ib have strong He absorption features, while SNe Ic have features of Si II but not He \citep{1985ApJ...296..379E, 1994ApJ...436L.135W}. The SNe IIb transition from SNe II to Ib, exhibiting hydrogen features at very early times that vanish during their later phases \citep{1997ARA&A..35..309F, 2017hsn..book..195G}. The SNe Ic-BL have very broad absorption features of Si II and are associated with long gamma-ray bursts \citep{2002ApJ...572L..61M}.
The loss of the outer stellar envelopes before the explosion of a SESN can be due to strong winds \citep[e.g., in Wolf-Rayet (WR) stars,][]{1995ApJ...448..315W, 2012A&A...542A..29G} or binary interactions \citep{2010ApJ...725..940Y}. 
The SNe IIn have narrow emission lines that arise from the presence of a dense H-rich circumstellar material \citep[CSM,][]{1990MNRAS.244..269S}. The same mechanism gives rise to SNe Ibn, with an He-rich CSM \citep{2008MNRAS.389..113P}, and the recently discovered SNe Icn, with a C- and O-rich CSM \citep[e.g.,][]{2021arXiv210807278F, 2022Natur.601..201G, 2023MNRAS.523.2530D, 2023A&A...673A..27N}. 
Superluminous (SL) SNe have absolute magnitudes $<-21$ mag \citep[e.g.,][]{2012Sci...337..927G, 2018SSRv..214...59M} and are classified into H-rich SLSNe-II \citep[e.g.,][]{2007ApJ...659L..13O} and H-poor SLSNe-I \citep{2011Natur.474..487Q}. 
Although SLSNe-II are thought to be similar to SNe IIn, the mechanism and progenitor stars behind the SLSNe-I are still not clear. 
Formation scenarios include interactions with an H-poor CSM \citep{2011ApJ...729L...6C, 2016ApJ...829...17S} or energy injections from a central compact object, possibly a magnetar \citep{2010ApJ...717..245K, 2010ApJ...719L.204W, 2013ApJ...770..128I, 2012MNRAS.426L..76D, 2013Natur.502..346N, 2016MNRAS.458.3455M, 2016Sci...351..257D, 2017ApJ...850...55N, 2023arXiv230903270H}.

 The rates at which each SN type occurs can help us understand their progenitor stars and explosion mechanisms. The dependence of the relative fractions of SN types and their rates on environmental properties can also reveal important differences between the progenitors of the various subtypes \citep[][]{2017ApJ...837..120G, 2019MNRAS.484.3785B, 2021MNRAS.500.5142F}. 
Recent large transient surveys, such as the Lick Observatory Supernova Search \citep[LOSS,][]{2000AIPC..522..103L, 2001ASPC..246..121F}, the Palomar Transient Factory \citep[PTF,][]{2009PASP..121.1395L}, and the Zwicky Transient Facility \citep[ZTF,][]{2019PASP..131a8002B}, have discovered thousands of SNe over the past decade. Dedicated spectroscopic classification surveys, such as PESSTO \citep{2015A&A...579A..40S} and the ZTF Bright Transient Survey \citep{2020ApJ...895...32F}, have enabled more reliable classifications of different transient types, largely increasing the known samples of different SN subtypes. These large SN samples, with high classification completeness, have allowed for very precise rate estimates for the different SN types.
\citet[][hereafter \citetalias{2011MNRAS.412.1441L}]{2011MNRAS.412.1473L, 2011MNRAS.412.1441L}, for instance, presented volumetric rates of nearby ccSNe for the largest sample to date, composed of 440 events discovered by the LOSS survey. A reanalysis of these results by \citet{2017ApJ...837..120G} also showed correlations of ccSN rates with galaxy properties, such as stellar mass and star-formation rate (SFR). LOSS was, however, a biased survey, targeting specific, generally more massive galaxies. Recently, \citet[][hereafter \citetalias{2020ApJ...904...35P}]{2020ApJ...904...35P} estimated ccSN rates using 313 SNe discovered by the ZTF, and \citet[][hereafter \citetalias{2021MNRAS.500.5142F}]{2021MNRAS.500.5142F} presented ccSN rates using 86 nearby SNe detected in the PTF survey. \citetalias{2021MNRAS.500.5142F} also presented rates for SESNe and SLSNe and compared these results with the SN host galaxy stellar mass. Other rate estimates for nearby SNe were made by \citet[][hereafter \citetalias{1999AA...351..459C}]{1999AA...351..459C}, \citet[][\citetalias{2014ApJ...792..135T}]{2014ApJ...792..135T}, \citet[][\citetalias{2015MNRAS.450..905G}]{2015MNRAS.450..905G}, \citet[][\citetalias{2015AA...584A..62C}]{2015AA...584A..62C}, and \citet{2025arXiv250404507M}.
The ccSN rates at higher redshifts have been estimated in several studies by: \citet[][hereafter \citetalias{2005AA...430...83C}]{2005AA...430...83C}, using nine spectroscopically classified ccSNe at $z = 0.26$; \citet[][hereafter \citetalias{2008AA...479...49B}]{2008AA...479...49B}, using 16 spectroscopically confirmed ccSNe from the Southern inTermediate Redshift ESO Supernova Search (STRESS, at $z = 0.2$); \citet[][hereafter \citetalias{2009AA...499..653B}]{2009AA...499..653B}, using 117 ccSNe from the Supernova Legacy Survey (SNLS, at $z = 0.3$); and \citet[][hereafter \citetalias{2012ApJ...757...70D}]{2012ApJ...757...70D} using 45 ccSNe from the Extended Hubble Space Telescope SN survey at $0.1 < z < 1.3$.

In this paper, we use the largest untargeted, magnitude-limited sample to date to report the volumetric rates and luminosity functions (LFs) of all ccSNe and their subtypes, including SESNe, interacting SNe, and SLSNe, from the All-Sky Automated Survey for Supernovae \citep[ASAS-SN,][]{2014ApJ...788...48S, 2017PASP..129j4502K, 2017MNRAS.467.1098H, 2017MNRAS.471.4966H, 2017MNRAS.464.2672H, 2019MNRAS.484.1899H, 2023MNRAS.520.4356N}. ASAS-SN is an all-sky untargeted survey that aims to monitor the entire sky to discover nearby and bright transient events. The survey began its operations in 2011, with the first alerts issued in 2013. In contrast to many previous surveys, ASAS-SN is unbiased with respect to the host galaxy properties. Additionally, the survey is nearly spectroscopically complete: due to its relatively low limiting magnitude ($m_{V} \sim 17.0$ mag), spectroscopic classifications have been obtained for almost all of the discovered events \citep[][]{2019MNRAS.484.1899H}. These properties make ASAS-SN ideal for population studies of nearby SNe.

The first paper of this series, \citet[][hereafter \citetalias{2023arXiv230611100D}]{2023arXiv230611100D}, reported the volumetric rates and LFs of SNe Ia discovered by ASAS-SN in the $V$ band, between 2014 and 2017. 
This is one of the most robust measurements of SNe Ia rates to date, due to spectroscopic completeness, sample size, and detailed completeness corrections.
In this work, we applied a similar methodology to estimate the volumetric rates and LFs of ccSNe. 
In Sect. \ref{sec:sample} we describe the sample and its properties. We describe the methodology used to compute the rates in Sect. \ref{sec:rate} and present the results in Sect. \ref{sec:res}. We summarize the results in Sect. \ref{sec:conc}.
Throughout this paper, we adopt a flat $\Lambda$-cold dark matter (CDM) cosmology with $H_0 = 70 \ \textrm{km s}^{-1} \textrm{Mpc}^{-1}$ and $\Omega_{m,0} = 0.3$.

\section{Sample selection}\label{sec:sample}

\begin{figure*}[t]
\centerline{
\includegraphics[trim=0.5cm 1cm 0.5cm 1cm, clip, scale=0.24]{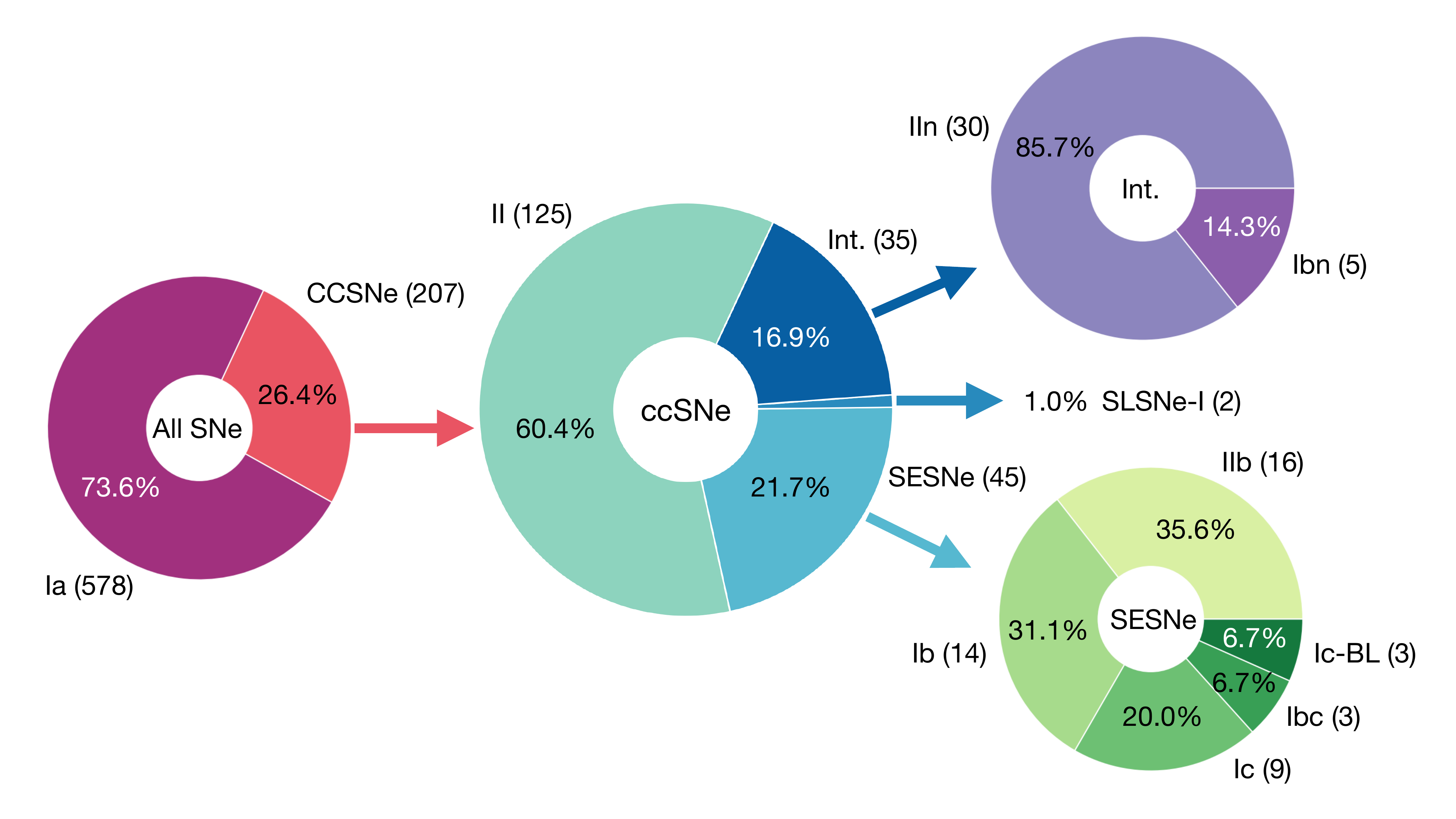}}
\caption{Fractions of all the SNe and their subtypes discovered or recovered by ASAS-SN between 2014 and 2017 \citep[][]{2017MNRAS.467.1098H, 2017MNRAS.471.4966H, 2017MNRAS.464.2672H, 2019MNRAS.484.1899H}. This sample is $97\%$ spectroscopically complete.  \label{fig:pie_chart}} 
\end{figure*}

\begin{figure}[t]
\centerline{
\includegraphics[scale=0.31]{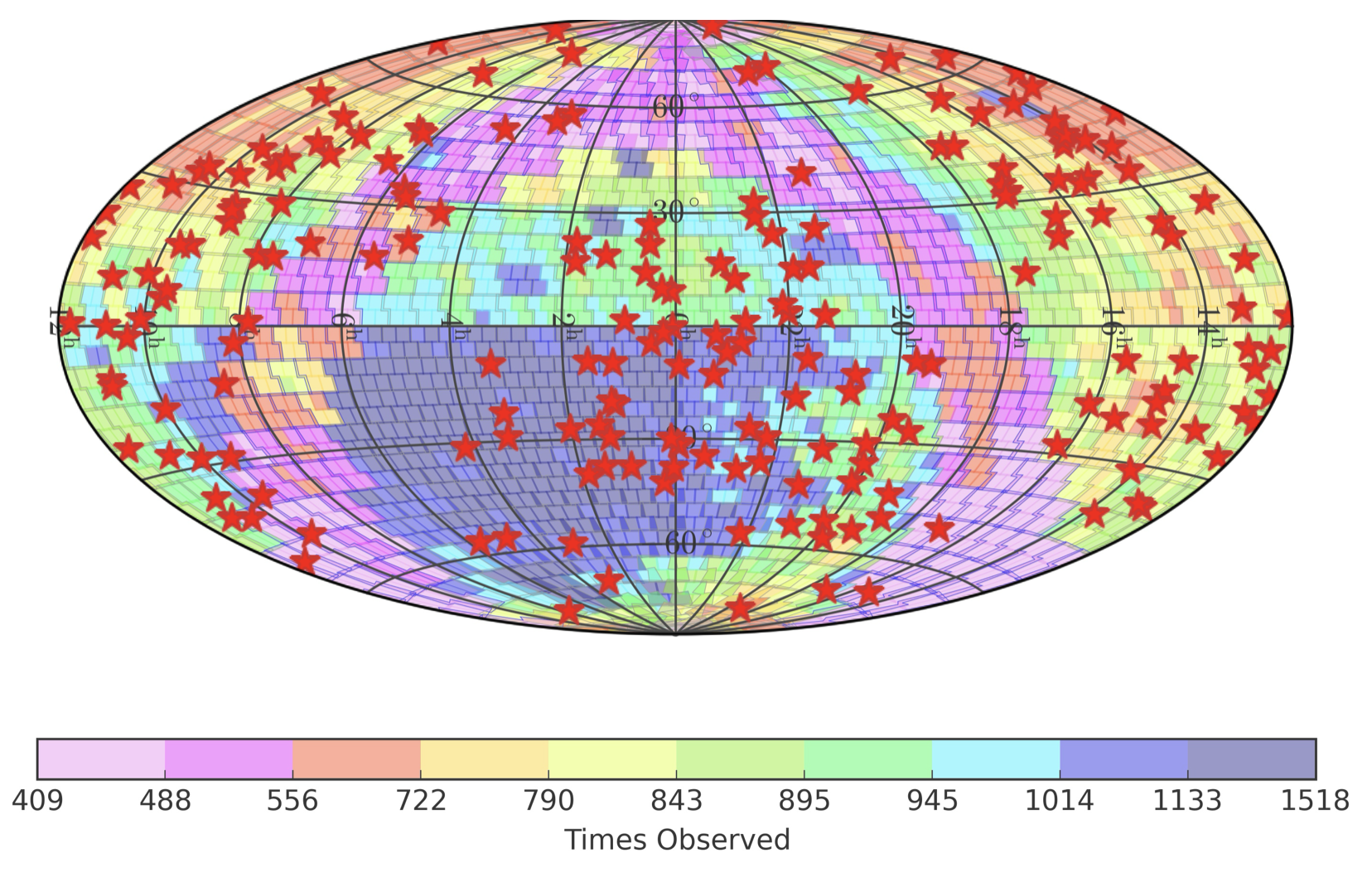}}
\caption{Equatorial coordinates of all the ccSNe discovered or recovered by ASAS-SN between 2014 and 2017 (red stars). The ccSNe are shown over the survey sky footprint in the same period. The color bar shows the number of images for individual observations in different areas of the sky. \label{fig:sky_map}} 
\end{figure}

The $V-$band ASAS-SN bright supernova catalogs \citep[][]{2017MNRAS.467.1098H, 2017MNRAS.471.4966H, 2017MNRAS.464.2672H, 2019MNRAS.484.1899H} report 1007 SNe observed between 2014 and 2017, of which 783 were discovered or recovered\footnote{By recovered, we mean an SN discovered first by another group that was later flagged in the ASAS-SN data.} by the survey. 
Due to the high spectroscopic completeness of ASAS-SN in this period, $97\%$ of the discovered events were spectroscopically classified \citep{2019MNRAS.484.1899H}. 
For this study, we double-checked the SN classifications and their associated spectra. We removed three events originally classified as ccSNe: PS16dtm, which is possibly a tidal-disruption event (TDE); ASASSN-17jz, an ambiguous nuclear event, possibly associated with an active galactic nucleus \citep{2022ApJ...933..196H}; and ASASSN-17qp, which was only found in the $g$ band. 

As shown in Fig. \ref{fig:pie_chart}, 578 ($73.6\%$) events are classified as SNe Ia, and 207 ($26.4\%$) as ccSNe in the ASAS-SN catalogs. 
Here, we analyze the 207 spectroscopically classified ccSNe found by ASAS-SN between UTC 2014-05-04.47 and UTC 2017-12-20.47. Of these, $60.4\%$ are SNe II, $21.7\%$ are SESNe, $16.9\%$ are interacting SNe, and two events ($1\%$) are SLSNe-I (although the nature of ASASSN-15lh is debated; see the discussion in Sect. \ref{sec:SLSN}).   
Among the SESNe, $35.6\%$ are SNe IIb, $31.1\%$ are SNe Ib, $20.0\%$ are SNe Ic, $6.7\%$ are SNe Ic-BL, and $6.7\%$ have ambiguous classifications and are labeled as SNe Ibc. 
For the interacting events, $85.7\%$ are SNe IIn, and $14.3\%$ are SNe Ibn.

Figure \ref{fig:sky_map} shows the distribution of the ccSNe in the sky, together with the number of observations of ASAS-SN.
Table \ref{tab:sn_prop} reports the properties of the ccSNe in our sample. The names, types, coordinates (RA and Dec), and host galaxy redshift ($z_{host}$) were retrieved from \citet[][]{2017MNRAS.467.1098H, 2017MNRAS.471.4966H, 2017MNRAS.464.2672H, 2019MNRAS.484.1899H}.
We estimated the $V-$band observed flux ($F_{V, peak}$), apparent magnitude ($m_{V,peak}$), and time at peak brightness ($t_{peak}$), by fitting the ccSN flux light curves with templates from \citet[][hereafter \citetalias{2019MNRAS.489.5802V}]{2019MNRAS.489.5802V}. 
We fit the observed light curve fluxes, as retrieved from the ASAS-SN Sky Patrol v2.0 \citep{2023arXiv230403791H}. 
The \citetalias{2019MNRAS.489.5802V} templates used are: SN2011ht, SN2009ip, SN2007pk, and SN2006aa (for SNe IIn); SN2004gt (for SNe Ic); SN2013ej, SN2004et, SN2014G, and ASASSN14jb (for SNe II); iPTF13bvn (for SNe Ib, Ibn, and Ibc); SN2008ax (for SNe IIb); and SN2009bb (for SNe Ic-BL).
The best fitting template is used to estimate $F_{V, peak}$, $m_{V,peak}$, and $t_{peak}$.
For the SLSNe-I Gaia17biu and ASASSN-15lh, we used Gaussian process regression to fit their light curves and estimate $m_{V,peak}$ and $t_{peak}$. We complemented our $V-$band photometry for Gaia17biu with data from \citet{2022CoSka..52a..46T} and \citet{2023ApJ...949...23Z}. 

The absolute $V-$band magnitudes

\begin{equation}\label{eq:MV}
    M_{V, peak} = m_{V, peak} - \mu (z) - A_{V,MW} - K_{V, peak}(z),
\end{equation}

\noindent depend on the distance modulus, $\mu (z)$, obtained using \texttt{astropy.cosmology} \citep{astropy:2013,astropy:2018}, the Milky Way extinction towards the SN, $A_{V,MW}$, taken from \citet{2011ApJ...737..103S}, and the $K-$correction, $K(z)$. The $K-$correction given by

\begin{equation}\label{eq:K}
    K_{V, peak}(z) = 2.5 \ \textrm{log} (1+z) \ + \ 2.5 \ \textrm{log} \left( \frac{\int F_{peak}(\lambda) S_V(\lambda) d\lambda}{\int F_{peak}(\lambda / (1 + z )) S_V(\lambda) d\lambda} \right),
\end{equation}

\noindent was obtained following \citet{1996PASP..108..190K}, where $S_V(\lambda)$ is the Bessel $V$ filter transmission \citep{1990PASP..102.1181B}, obtained with \texttt{speclite}\footnote{\url{https://speclite.readthedocs.io/en/latest/}}, and $F_{peak}(\lambda)$ is the observed SN spectrum at peak brightness. 
We do not apply corrections for host-galaxy extinction here.
The values of $M_{V, peak}$, $A_{V,MW}$, and $K_{V, peak}$ are reported in Table \ref{tab:sn_prop}.
For each ccSN subtype, we used the \citetalias{2019MNRAS.489.5802V} spectra at peak brightness associated with the photometric templates. 
For SNe Ibn and SLSNe, we used their observed spectra around peak brightness (ASASSN-14ms, \citeauthor{2018MNRAS.475.2344V} \citeyear{2018MNRAS.475.2344V}; ASASSN-15ed, \citeauthor{2015MNRAS.453.3649P} \citeyear{2015MNRAS.453.3649P}; SN2015U, \citeauthor{2015MNRAS.454.4293P} \citeyear{2015MNRAS.454.4293P}; Gaia17biu, \citeauthor{2018ApJ...853...57B} \citeyear{2018ApJ...853...57B}; ASASSN-15lh: \citeauthor{2016Sci...351..257D} \citeyear{2016Sci...351..257D}), except for ASASSN-14dd and ASASSN-17gi, for which we used the spectrum of SN2006jc at peak \citep{2007Natur.447..829P}.

\begin{figure}[t]
\centerline{
\includegraphics[scale=0.62]{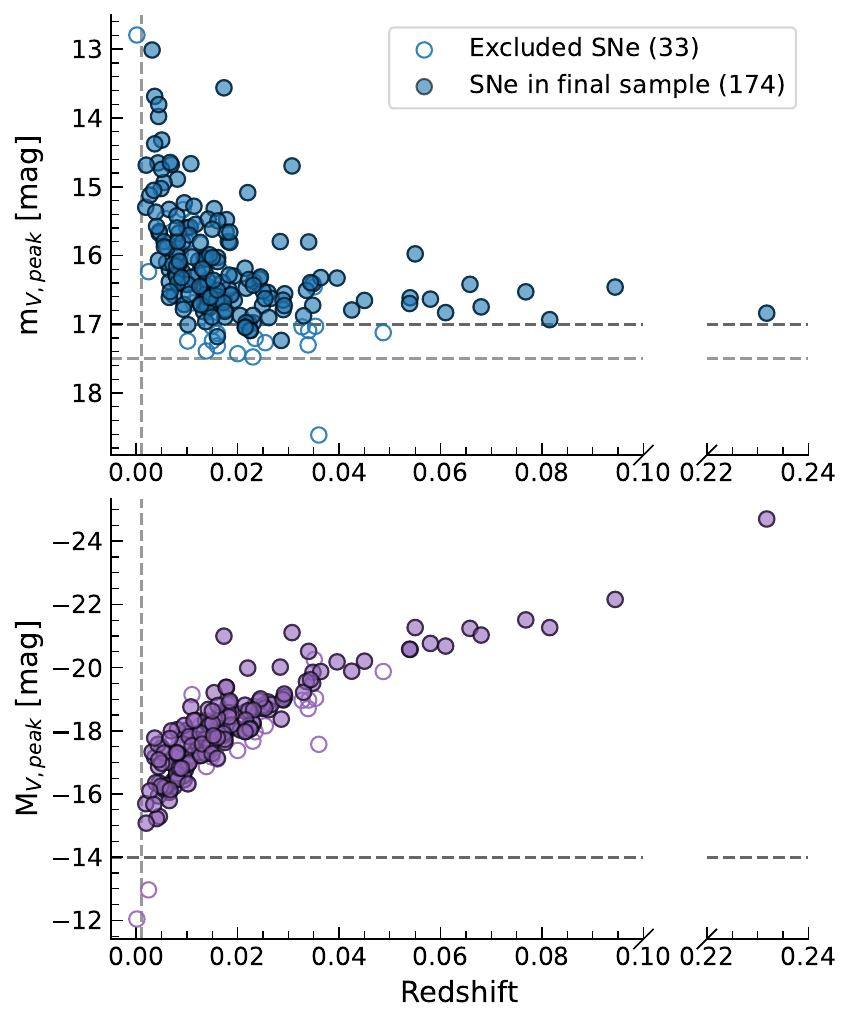}}
\caption{Distribution of apparent (top) and absolute magnitude (bottom) at peak brightness for the 207 events in the initial sample. We restrict the sample to events with $m_{V, peak} \leq 17.0$ and $17.5$ mag, $M_{V, peak} \leq -14.0$ mag, and $z_{min} \geq 0.001$ (dashed lines). The open circles correspond to the excluded 33 SNe, and the filled circles correspond to the 174 SNe in the final sample. \label{fig:mag_z}} 
\end{figure}

Figure \ref{fig:mag_z} shows the distributions of $m_{V, peak}$ and $M_{V, peak}$ for the 207 ccSNe. 
The events span $0.000133 < z < 0.2318$, with a median of $z_{med} = 0.014$. 
The brightest event has $m_{V, peak} = 12.79$ mag, and few have $m_{V, peak} > 17.0$ mag. The sample has a median peak apparent magnitude of $16.37$ mag and a median peak absolute magnitude of $-17.82$ mag.

For our rate estimates, we used events with $m_{V, peak} < 17.0$ mag, corresponding to the limiting magnitude of ASAS-SN, except for events with long-duration plateaus in their light curves, where we considered events with $m_{V, peak} < 17.5$ mag.
We excluded events with a Galactic latitude of $|b| < 15^\circ$, to avoid highly extincted regions in the Galaxy. 
We also limited the absolute magnitude to $M_{V, peak} \leq -14.0$ and the redshift range to $0.001 < z < 0.1$ (we estimated the rate for ASASSN-15lh separately; see the discussion in Sect. \ref{sec:SLSN}). 
We used $B-$band Tully-Fisher distances \citep{1988cng..book.....T} for all events in the redshift range $0.001 < z < 0.005$. 
These cuts result in a final sample with 174 events (97 II, 29 IIn, 13 Ib, 12 IIb, 7 Ic, 3 Ib/c, 3 Ibn, 2 Ic-BL, and 2 SLSNe-I). 

%----------------

\begin{table*}[t!]
\renewcommand{\arraystretch}{1.1}
\small
\caption{General properties of the SNe. \label{tab:sn_prop}}     
\centering          
\begin{tabular}{lccclcccclccc}  
\hline       
Name & Type$^a$ & RA& Dec & $z_{host}$ & $m_{V, peak}$ & $M_{V, peak}$ & $F_{V, peak}$ & $t_{peak}$  & $\mu(z)$ & $A_{V,MW} \ ^b$ & $K_{V, peak}$\\ 
 & & [J2000] & [J2000] & & [mag] & [mag] & [mJy] & [JD] & [mag] & [mag] & [mag] \\ 
\hline   
ASASSN-14at & II & 17 55 05.31 & +18 15 27.4 & 0.010431 & 16.56 & -16.99 & 0.853 & 56782.40 & 33.27 & 0.26 & 0.030 \\
ASASSN-14az & IIb & 23 44 48.05 & -02 07 02.1 & 0.0067 & 14.65 & -17.76 & 4.976 & 56793.75 & 32.30 & 0.10 & 0.009 \\
ASASSN-14bu & II & 11 18 41.03 & +25 09 59.9 & 0.0254 & 16.63 & -18.72 & 0.805 & 56812.40 & 35.22 & 0.05 & 0.074 \\
ASASSN-14dp & II & 11 21 58.33 & -37 54 24.6 & 0.009159 & 15.81 & -17.59 & 1.711 & 56866.00 & 32.98 & 0.38 & 0.040 \\
ASASSN-14dq & II & 21 57 59.97 & +24 16 08.1 & 0.010424 & 15.70 & -17.82 & 1.885 & 56846.30 & 33.27 & 0.22 & 0.037 \\
ASASSN-14il & IIn & 00 45 32.55 & -14 15 34.6 & 0.021989 & 15.09 & -19.98 & 3.323 & 56962.85 & 34.91 & 0.07 & 0.096 \\
... \\
\hline                  
\end{tabular}
\tablefoot{The entire table is available in the online journal. $^a$ All the supernova types, coordinates, and host galaxy redshifts were obtained from \citet[][]{2017MNRAS.467.1098H, 2017MNRAS.471.4966H, 2017MNRAS.464.2672H, 2019MNRAS.484.1899H}. $^b$ The Galactic extinction, $A_{v,MW}$, was obtained through the NASA-IPAC Extragalactic Database (NED).}
\end{table*}

\section{Rate computations} \label{sec:rate}

\begin{figure*}[t]
    \centerline{
    \includegraphics[scale=0.58]{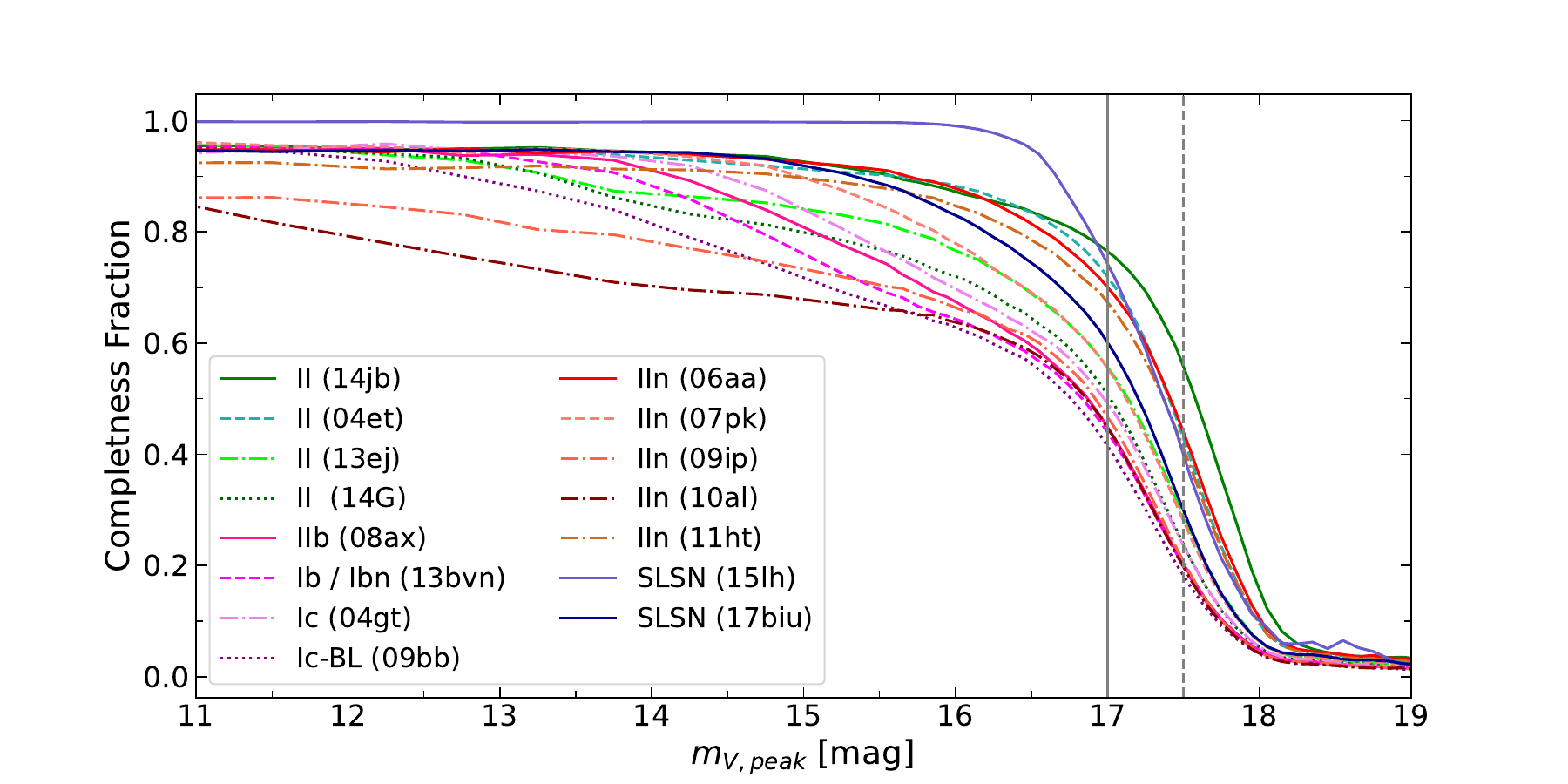}}
    \caption{Completeness fractions for different ccSN templates as a function of peak apparent magnitude ($m_{V, peak}$). 
    The vertical solid line shows the limiting magnitude of the analysis at $m_{V} = 17.0$ mag, while the vertical dashed line shows the limit used for light curves with long-duration plateaus at $m_{V} = 17.5$ mag. \label{fig:completeness}} 
    \end{figure*}

    \begin{figure}[t]
    \centering
    \includegraphics[width=0.46\textwidth]{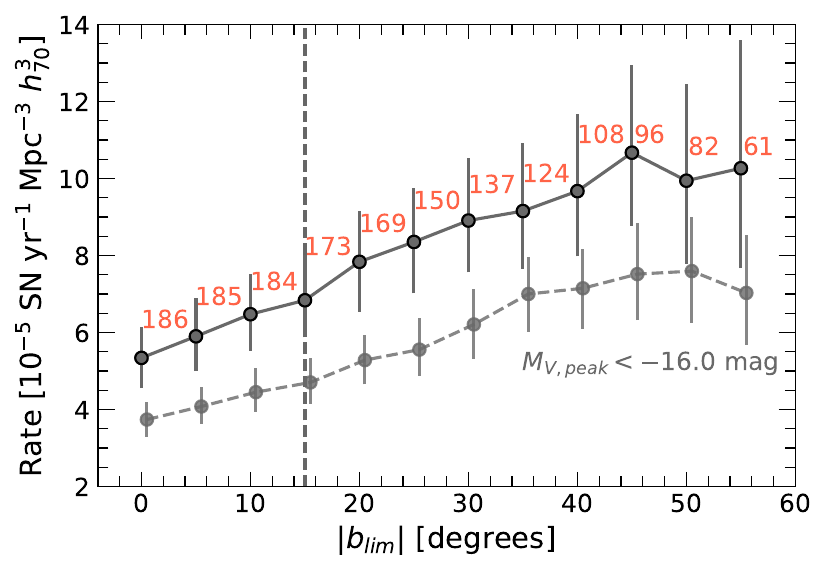}
    \includegraphics[width=0.44\textwidth]{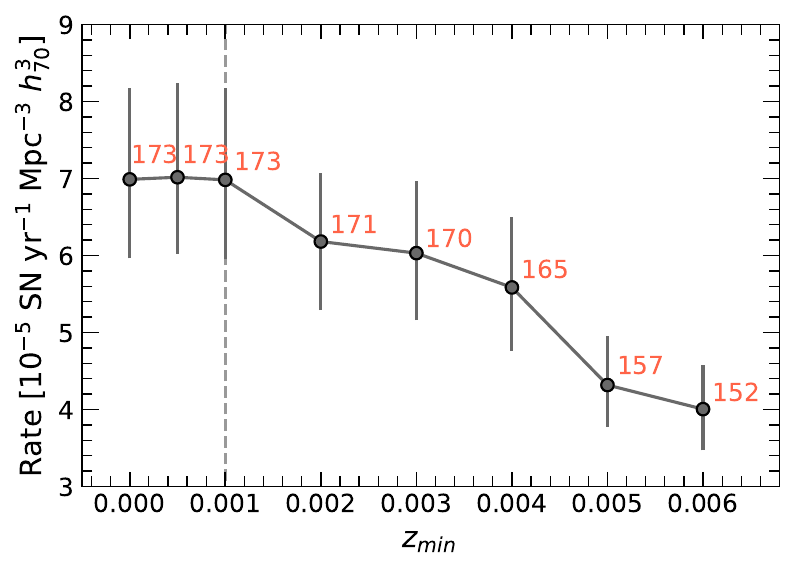}
    \includegraphics[width=0.455\textwidth]{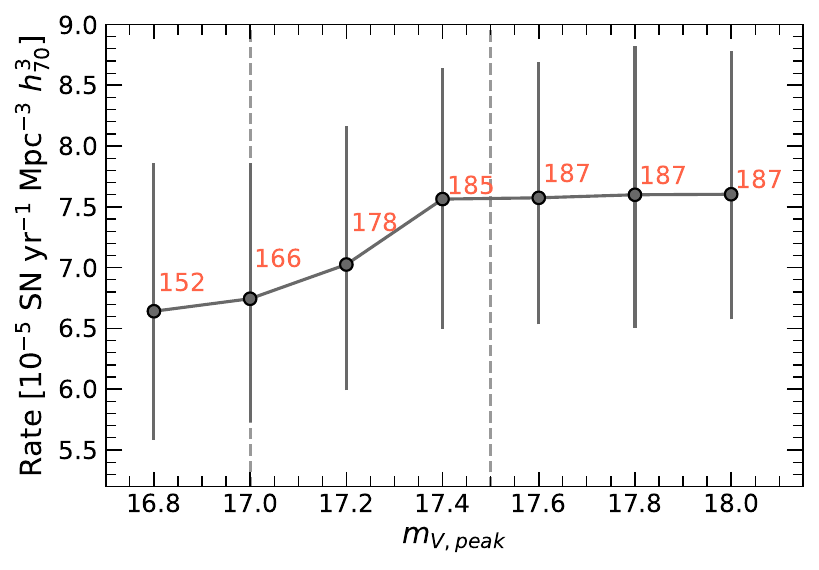}
        \caption{ccSN volumetric rate as a function of the minimum Galactic latitude ($b_{lim}$, top panel), minimum redshift ($z_{min}$, middle panel), and maximum apparent magnitude at peak ($m_{V,peak}$, bottom panel). 
        The number of ccSNe included in each subsample is given in red, while the vertical dashed lines show the limits used for the final sample.  \label{fig:rate_vs_params}} 
    \end{figure}

    \begin{figure*}[t]
    \centerline{
    \includegraphics[scale=0.54]{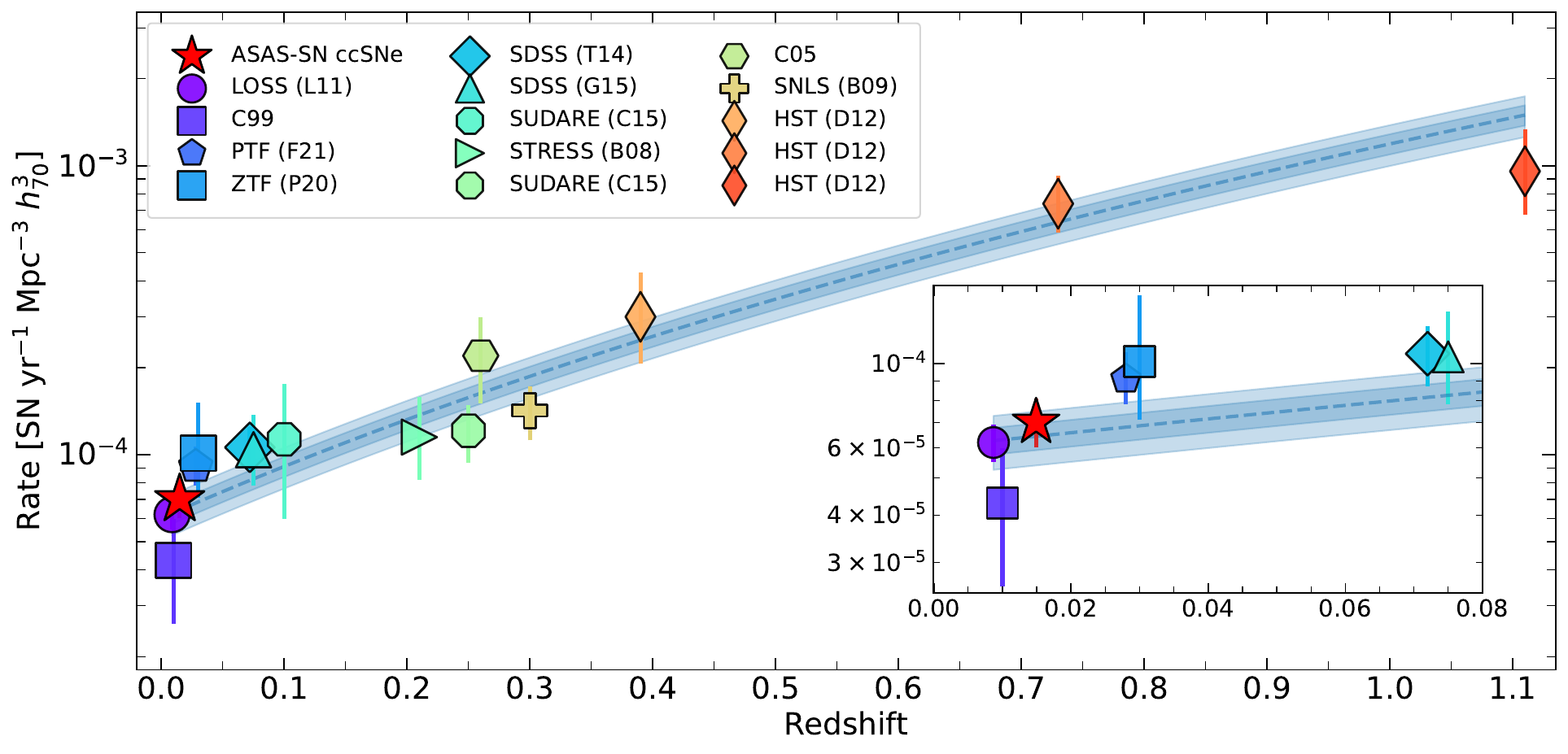}}
    \caption{Volumetric ccSN rates as a function of redshift. We compare our ccSN rate (red star) estimate with \citetalias{1999AA...351..459C}, \citetalias[][]{2005AA...430...83C}, \citetalias[][]{2008AA...479...49B}, \citetalias[][]{2009AA...499..653B}, \citetalias[][]{2011MNRAS.412.1441L}, \citetalias[][]{2012ApJ...757...70D}, \citetalias[][]{2014ApJ...792..135T}, \citetalias[][]{2015MNRAS.450..905G}, \citetalias[][]{2015AA...584A..62C}, \citetalias{2020ApJ...904...35P}, and \citetalias[][]{2021MNRAS.500.5142F}. The dashed blue line is an SFH model scaled to these ccSN rates, following a power law with $\beta = 4.3$, with the dark and light blue regions showing the one and two $\sigma$ uncertainties of the model, respectively. 
    The inset zooms in on the ccSN rates for the nearby universe ($0.00 < z < 0.08$).\label{fig:rate_z}} 
    \end{figure*}

We followed a similar procedure used by \citetalias{2023arXiv230611100D} to estimate the volumetric rates of ccSNe detected by the ASAS-SN survey; see \citetalias{2023arXiv230611100D} for a more in-depth discussion of the complete methodology.
The probability of detection ($p$) of an event in a single ASAS-SN epoch is only dependent on the signal-to-noise ratio ($S/N$) of the observation. It is well described by the empirical relation

\begin{equation} \label{eq:probability}
    p = \begin{cases} 
      0 & \textrm{for} \ S/N \leq 5\\
      0.65 \left( 1 - \frac{12 - S/N}{7} \right) & \textrm{for} \ 5 \leq S/N \leq 12 \\
      0.65 & \textrm{for} \ S/N \geq 12, 
   \end{cases}
\end{equation}

\noindent estimated from all supernova observations over the survey period (discovered, recovered, and missed). 

We drew $N_{LC} = 164,191$ random light curves at different sky positions to sample the noise and cadence properties of the survey.  For each SN template, we randomly selected redshifts $z$ (following the comoving volume within the range $0<z < z_{lim}$, where $z_{lim}$ set by the peak absolute magnitude) and peak times, $t_{peak}$. We then added the implied fluxes to a randomly selected light curve and used Eq. \ref{eq:probability} to determine whether the trial would be detected. We did this for $M= 100 \ N_{LC}$ trials as a function of peak apparent magnitude, leading to a detection fraction of $F_1=N/M$, where $N$ corresponds to the number of detections. 

Figure \ref{fig:completeness} shows the resulting completeness functions for the different ccSN templates. The primary driver of the differences is the time spent near peak. SNe~II with long plateaus (e.g., ASASSN-14jb-like and SN2004et-like) or SNe IIn with long duration light curves (e.g., SN2006aa-like and SN2011ht-like) have higher completeness at a given peak magnitude than those with shorter plateaus (e.g., SN2014G-like SNe II, SN2008ax-like SNe IIb, and SN2010al-like SNe IIn).

A second completeness factor was used to correct for differences in volume between the redshift intervals, since the simulations adjust $z_{lim}$ with the peak absolute magnitude. With $z_{max} = 0.1$, we obtain $F_2(M_{V,peak}) = V(z_{lim}(M_{V,peak})) / V(z_{max})$ for a comoving volume $V(z)$. Finally, the statistical weight for the $i^{\textrm{th}}$ observed SN is $w_i = (F_{1,i}F_{2,i})^{-1}$, and the volumetric rate of ccSNe, per unit comoving volume and time, is 

\begin{equation}\label{eq:rate}
  R = \frac{\Sigma^{N}_{i=1} w_i}{V \ \Delta t \ (1 - \textrm{sin} \ b_{lim} )},  
\end{equation}

\noindent where $V = \frac{4}{3} \pi \left( d^3_{max} - d^3_{min} \right)$ is the total comoving volume corresponding to $0.001 \leq z \leq 0.1$, $\Delta t = 4.0$~yr is the time interval between UTC 2014-01-01 and UTC 2017-12-31, and $1 - \textrm{sin} \ b_{lim}$ is the volume fraction with Galactic latitude $|b_{lim}| \geq 15^\circ$.

\section{Results and discussions} \label{sec:res}

\subsection{Volumetric ccSN rates} \label{subsec:ccSN_rates}

    \begin{figure}[t]
    \centerline{
    \includegraphics[trim=9cm 2cm 6cm 0cm, clip, scale=0.46]{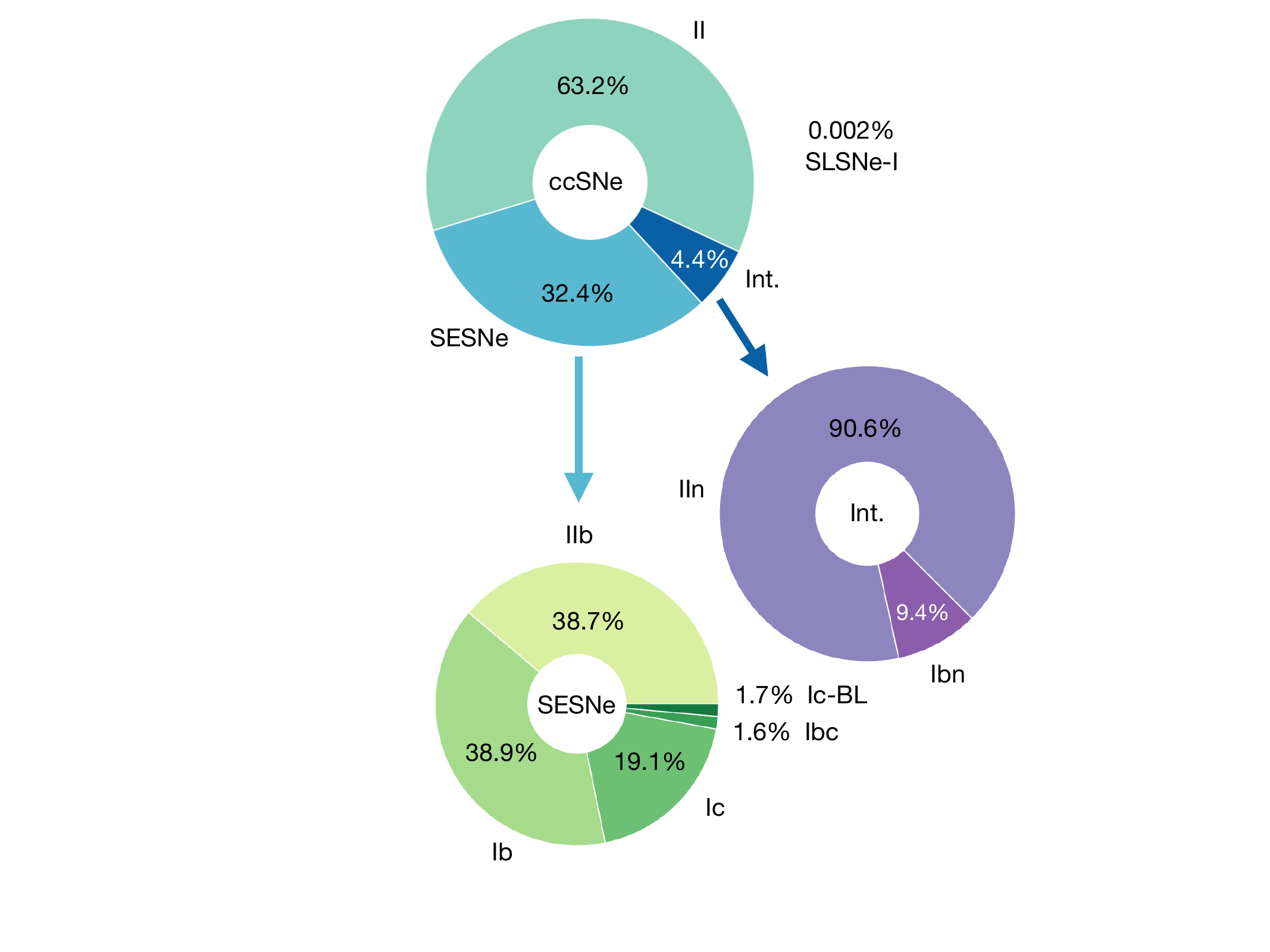}}
    \caption{Fractions of the volumetric rates associated with different ccSN subtypes. Fractions are shown for SNe II, SESNe and its subtypes (SNe Ib, Ic, IIb, Ibc, and Ic-BL), and interacting SNe (Int.) and its subtypes (SNe IIn and Ibn). The fractions relative to the total ccSN rate are reported in Table \ref{tab:ccSN_rates_sub}. \label{fig:pie_plot_vol}} 
    \end{figure}

    \begin{table}
    \renewcommand{\arraystretch}{1.3}
    \small
    \caption{Volumetric ccSN rates for varying peak absolute magnitude ranges. \label{tab:rates_MVmin}}     
    \centering          
    \begin{tabular}{cccc}  
    \hline       
    $M_{V, peak, \textrm{range}}$ & $N_{\textrm{ccSN}}$ & $R_{\textrm{ccSN}}$ & $\sigma_R$ / $R$\\ 
     & & [$10^{-5} \ \textrm{yr}^{-1} \ \textrm{Mpc}^{-3} \ h^{3}_{70}$] & \\ 
    \hline                    
    [-15.0, -22.0] & 173 & $6.9^{+1.1}_{-0.9}$ & $^{+16\%}_{-14\%}$ \\
    
    [-15.5, -22.0] & 170 & $5.6^{+0.7}_{-0.7}$ & $^{+13\%}_{-12\%}$ \\
    
    [-16.0, -22.0] & 166 & $4.8^{+0.6}_{-0.6}$ & $^{+12\%}_{-12\%}$ \\
    \hline                  
    \end{tabular}
    \end{table}

    \begin{table}
    \renewcommand{\arraystretch}{1.3}
    \small
    \caption{Volumetric ccSN rates for varying redshift ranges. \label{tab:rates_zmin}}     
    \centering          
    \begin{tabular}{cccc}  
    \hline       
    $z_{\textrm{range}}$ & $N_{\textrm{ccSN}}$ & $R_{\textrm{ccSN}}$ & $\sigma_R$ / $R$\\ 
     & & [$10^{-5} \ \textrm{yr}^{-1} \ \textrm{Mpc}^{-3} \ h^{3}_{70}$] & \\ 
    \hline                    
    [0.001, 0.1] & 173 & $7.0^{+1.0}_{-0.9}$ & $^{+15\%}_{-14\%}$ \\
    
    [0.003, 0.1] & 170 & $6.0^{+0.9}_{-0.8}$ & $^{+16\%}_{-13\%}$ \\
    
    [0.005, 0.1] & 157 & $4.3^{+0.5}_{-0.5}$ & $^{+13\%}_{-12\%}$ \\
    \hline                  
    \end{tabular}
    \end{table}

        %%%%%%%%%%%%%%%%
    
    \begin{table}[t!]
    \renewcommand{\arraystretch}{1.3}
    \small
    \caption{Volumetric ccSN rates. \label{tab:ccSN_rates}}     
    \centering          
    \begin{tabular}{lcccl}  
    \hline       
    $z_{med}$ & N$_{\textrm{ccSN}}$ & Rate & Survey / Ref. \\ 
    &  & [$\times 10^{-5}$ yr$^{-1}$ Mpc$^{-3}$ $h^{3}_{70}$] & \\
    \hline        
    0.0149 & 173 & \textbf{$7.0^{+1.0}_{-0.9}$} & ASAS-SN $^\dagger$\\
    0.0087 & 440 & $6.2\pm0.7$ & LOSS (\citetalias{2011MNRAS.412.1441L}) \\
    0.01 & 67.4  & $4.3\pm1.7$ & \citetalias{1999AA...351..459C} $^a$\\ 
    0.028 & 86 & $9.1^{+1.6}_{-1.3}$ & PTF (\citetalias{2021MNRAS.500.5142F}) \\ 
    0.03 & 313 & $10.1^{+5.0}_{-3.0}$ & ZTF (\citetalias{2020ApJ...904...35P}) \\
    0.072 & 89 & $10.6 \pm1.9$ & SDSS (\citetalias{2014ApJ...792..135T}) \\ 
    0.075 & 16 & $10.4^{+3.3}_{-2.6}$ & SDSS (\citetalias[][]{2015MNRAS.450..905G}) \\ 
    0.10 & 5.9 & $11.3^{+6.2}_{-5.3}$ & SUDARE (\citetalias[][]{2015AA...584A..62C}) \\ 
    0.21 & 16 & $11.5^{+4.3}_{-3.3}$ & STRESS (\citetalias{2008AA...479...49B}) $^a$\\
    0.25 & 26.2 & $12.1^{+2.7}_{-2.7}$ & SUDARE (\citetalias[][]{2015AA...584A..62C})\\
    0.26 & 18.8 & $22.0^{+8.0}_{-7.0}$ & \citetalias{2005AA...430...83C} $^a$\\
    0.30 & 117 & $14.2 \pm3.0$ & SNLS (\citetalias{2009AA...499..653B})\\
    0.39 & 9 & $30.0^{+12.8}_{-9.4}$ & HST (\citetalias{2012ApJ...757...70D}) $^b$\\
    0.73 & 25 & $73.9^{+18.6}_{-15.2}$ & HST (\citetalias{2012ApJ...757...70D})\\
    1.11 & 11 & $95.7^{+37.6}_{-28.0}$ & HST (\citetalias{2012ApJ...757...70D})\\
    \hline                  
    \end{tabular}
    \tablefoot{ $^\dagger$ This work. a. Rates scaled to our assumed cosmology; 
    b. Values that apply corrections for extinction / missing SNe.}
    \end{table}

    %%%%%%%%%%%%%%%%

    The total ccSN volumetric rate for our sample, limited to absolute magnitudes $M_{V,peak} \leq -14$ mag and a redshift range of $0.001 < z < 0.1$ with a median redshift of $z_{\textrm{med}} = 0.0149$, is 

    $$ R_{\textrm{ccSN}} = 7.0^{+1.0}_{-0.9} \times 10^{-5} \ \textrm{yr}^{-1} \ \textrm{Mpc}^{-3} \ h^{3}_{70}, $$

    \noindent with the uncertainties given by the one $\sigma$ range of $10^3$ bootstrap resamplings of the rate calculation. 
This estimate takes into account the SLSN-I Gaia17biu but does not include ASASSN-15lh.
Table \ref{tab:rates_MVmin} gives the dependence of the rate on the minimum absolute magnitude at peak, $M_{V,peak}$. The rate significantly decreases with increasing minimum luminosity, because the rarer, low-luminosity SNe disproportionately contribute to the correction weights due to their large completeness corrections.
Although the rates with lower $M_{V,peak}$ have fractional uncertainties ($\sigma_R / R$), they represent only lower limits for the total ccSN rate, as a non-negligible number of low luminosity events are excluded.
Figure \ref{fig:rate_vs_params} shows the variation of the estimated ccSN volumetric rates as a function of $b_{lim}$, $z_{min}$, and the limiting $m_{V, peak}$.
There is significant variation in the volumetric rate with $b_{lim}$, with a flattening observed after $b_{lim} > 45^\circ$, as shown in the top panel. 
This is mostly due to lower-luminosity SNe being preferentially detected at higher latitudes, which generates larger completeness corrections with increasing $b_{lim}$.
This effect is also indicated by the dashed gray curve, which shows the dependence of the rate on $b_{lim}$ for events with $M_{V,peak} < -16.0$ mag. When the lower-luminosity SNe are excluded, the variation in the rates with $b_{lim}$ is lower, making the curve flatten. 
The decreasing rates with increasing $z_{min}$ (middle panel, for a $z_{min}$ range of $0.001 - 0.005$) is a similar effect, as the loss of low-luminosity SNe significantly affects the rate estimate for larger minimum redshifts.
Table \ref{tab:rates_zmin} reports the rates for three different minimum redshifts and shows that the fractional uncertainty decreases for larger $z_{min}$, similar to the trend seen in Table \ref{tab:rates_MVmin}.
Finally, there is only a small variation in the volumetric rates with the limiting apparent magnitude at peak brightness (bottom panel, for $m_{V, \ peak} = 16.8 - 18.0$ mag), demonstrating the consistency of our completeness corrections.
In Table \ref{tab:ccSN_rates} and Fig. \ref{fig:rate_z}, we compare our rate with those from \citetalias{1999AA...351..459C}, \citetalias[][]{2005AA...430...83C}, \citetalias[,][]{2008AA...479...49B}, \citetalias[,][]{2009AA...499..653B}, \citetalias[,][]{2011MNRAS.412.1441L}, \citetalias[][]{2012ApJ...757...70D}, \citetalias[][]{2014ApJ...792..135T}, \citetalias[][]{2015MNRAS.450..905G}, \citetalias[][]{2015AA...584A..62C}, \citetalias{2020ApJ...904...35P} and \citetalias[][]{2021MNRAS.500.5142F}. 
Our value is consistent with the previous estimate of the nearby ccSN rate of \citetalias[][]{2011MNRAS.412.1441L} (within 1 $\sigma$)
    and with that of \citetalias{1999AA...351..459C} (within $1.3 \ \sigma$). 
An important difference with the estimates of \citetalias[][]{2011MNRAS.412.1441L} and \citetalias{1999AA...351..459C} is that they both used galaxy-targeted samples (instead of the all-sky untargeted search of ASAS-SN), which introduces systematics especially when estimating the dependence of rates on galaxy properties (see Sect. \ref{sec:galaxy}).
Our result is also consistent with the rate estimates at relatively larger median redshifts from \citetalias{2021MNRAS.500.5142F} (within $1.3 \ \sigma$)
and \citetalias{2020ApJ...904...35P} (within 1 $\sigma$).
The dashed blue line in Fig. \ref{fig:rate_z} shows a star-formation history (SFH) model in the form $\rho = \rho_0 (1+z)^{\beta}$, with the parametrization $\beta = 4.3$ from \citet{2011ApJ...730...61K}, and $\rho_0 = (6.04 \pm 0.48) \times 10^{-5} \ \textrm{yr}^{-1} \ \textrm{Mpc}^{-3} \ h^{3}_{70}$, scaled to fit the observed values. 
{Our result is consistent with the power-law coefficient of SFR data, as expected due to the short lifespan of the ccSN progenitor stars (although other studies suggest different coefficients for the nearby universe). 
We used the best-fitting model of \citet{2014ARA&A..52..415M} to obtain an SFR density of $\sim 0.015 \ \textrm{M}_\odot \ \textrm{yr}^{-1} \ \textrm{Mpc}^{-3} $ at $z = 0.0149$ and estimate the ratio of ccSN\ rate to SFR of $\sim 0.0045 \ \textrm{M}_\odot^{-1}$. A more detailed exploration of the normalization against SFR estimates should be performed in a future work in this series.}

 In Table \ref{tab:ccSN_rates_sub}, we report the volumetric rates for all the ccSN subtypes in our sample. These values were estimated by applying the same methodology described in Sect. \ref{sec:rate} to each classification subsample. The uncertainties are also given by the one $\sigma$ results of $10^3$ bootstrap runs. 
In Fig. \ref{fig:pie_plot_vol} and Table \ref{tab:ccSN_rates_sub}, we show the fractions of the total volumetric rates for the different ccSN subtypes.
We find a volumetric rate of SNe II of $(4.3\pm{0.8}) \times 10^{-5} \ \textrm{yr}^{-1} \ \textrm{Mpc}^{-3} \ h^{3}_{70}$, corresponding to $63.2\%^{+14.4\%}_{-13.8\%}$ of the occurrence rates of ccSNe.
This value is consistent with the estimate from the LOSS survey (within 1 $\sigma$) of ($4.4\pm{1.4} \times 10^{-5}) \ \textrm{yr}^{-1} \ \textrm{Mpc}^{-3}$ by \citetalias[][]{2011MNRAS.412.1441L}. 
Recently, \citet{2025arXiv250219493D} reported a volumetric rate of SNe IIP of $(3.9\pm{0.4}) \times 10^{-5} \ \textrm{yr}^{-1} \ \textrm{Mpc}^{-3}$, using 330 events discovered by the ZTF. We used the 34 SNe II exhibiting long-duration plateaus in their light curves (the events fit with the light curve templates of ASASSN-14jb and SN2004et) to estimate a volumetric rate of $2.1^{+0.8}_{-0.5} \times 10^{-5} \ \textrm{yr}^{-1} \ \textrm{Mpc}^{-3} \ h^{3}_{70}$. This value is $\sim 1.7$ times lower and shows a $\sim 2.4 \sigma$ difference from that found by \citet{2025arXiv250219493D}.
The SESN rate for our sample is

$$ R_{\textrm{SESN}} = 2.2^{+0.8}_{-0.6} \times 10^{-5} \ \textrm{yr}^{-1} \ \textrm{Mpc}^{-3} \ h^{3}_{70}, $$

\noindent using the 37 events classified as SNe IIb, Ib, Ic, Ib/c, and Ic-BL. Our result is consistent (within 1 $\sigma$) with the estimate by \citetalias{2021MNRAS.500.5142F} ($2.41^{+0.64}_{-0.81} \times 10^{-5} \ \textrm{yr}^{-1} \ \textrm{Mpc}^{-3}$), from a sample of 24 SESNe with $z_{med} = 0.028$.
The volumetric rate of SNe II is $\sim 1.9$ times higher than the rate of SESNe ($32.3\%^{+12.2\%}_{-9.4\%}$ of the total ccSN rate), and $\sim 14.3$ times higher than the rate of SNe IIn ($3.0^{+1.2}_{-1.0} \times 10^{-6} \ \textrm{yr}^{-1} \ \textrm{Mpc}^{-3} \ h^{3}_{70}$). 
Our occurrence fraction of SNe IIn, of $4.4\%^{+1.5\%}_{-1.5\%}$, is consistent with the estimate of $\sim 4.7\%$ by \citet{2023A&A...670A..48C}, while our estimated fraction for SNe Ibn, of $0.45\%^{+0.44\%}_{-0.42\%}$, is lower than the $\sim 1.0\%$ estimated by \citet{2022ApJ...927...25M}.
SNe IIb and Ib have similar volumetric rates, 
that are $\sim 2$ times higher than the volumetric rate of SNe Ic. 
This is similar to the ratio found by \citet{2017PASP..129e4201S}, $R_{Ib} / R_{Ic} = 2.7\pm0.9$.

Type Ic-BL SNe are associated with the afterglow of long-duration gamma-ray bursts \citep[LGRBs, e.g.,][]{1999ApJ...516..788W, 2002ApJ...572L..61M}. Their rates can, therefore, provide important constraints on the occurrence and nature of LGRBs \citep[e.g.,][]{2016A&A...587A..40P}.
From the two SNe Ic-BL discovered by ASAS-SN, we estimate a volumetric rate of $R_{Ic-BL} = 3.5^{+3.8}_{-2.2} \times 10^{-7} \ \textrm{yr}^{-1} \ \textrm{Mpc}^{-3} \ h^{3}_{70}$.
The combined volumetric rate of SNe Ic and Ic-BL is $R_{Ic + Ic-BL} = 4.4^{+2.8}_{-3.5} \times 10^{-6} \ \textrm{yr}^{-1} \ \textrm{Mpc}^{-3} \ h^{3}_{70}$, and the combined volumetric rate considering the SNe classified as Ibc is $R_{Ic + Ic-BL + Ibc} = 4.7^{+2.8}_{-3.5} \times 10^{-6} \ \textrm{yr}^{-1} \ \textrm{Mpc}^{-3} \ h^{3}_{70}$.

The total volumetric ccSN rate reported here is $\sim 3.0$ times higher than the total volumetric SN Ia rate of $(2.28\pm0.20) \times 10^{-5} \ \textrm{yr}^{-1} \ \textrm{Mpc}^{-3} \ h^{3}_{70}$, estimated for the ASAS-SN survey in \citetalias{2023arXiv230611100D} (obtained using 404 SNe Ia for a $z_{\textrm{med}} = 0.024$). 
This value is higher than the ratio of $R_{ccSN} / R_{Ia} \approx 2.0$ in \citetalias[][]{2011MNRAS.412.1441L}, but slightly lower than the $R_{ccSN} / R_{Ia} \approx 4.2$ in \citetalias{2021MNRAS.500.5142F}.

    Due to the lack of complete multiband photometry in our sample, we did not attempt to perform any host galaxy extinction corrections. 
    \citet{2012ApJ...756..111M} suggested that a large fraction of ccSNe might be subject to strong levels of extinction due to their host galaxy dust. They estimated that $\sim 20\%$ of ccSNe in the local Universe are missed by optical surveys, while \citet{2019ApJ...886...40J} reported an even higher fraction of $\sim 38\%$ of hidden ccSNe \citep[also see][]{2021MNRAS.506.4199F}.
    These obscured ccSNe could explain the mismatch between the local SFR and estimates of the SFR from ccSN rates \citep[e.g.,][]{2006ApJ...651..142H, 2011ApJ...738..154H, 2013ApJ...769..113H}. 
    If the large fractions of missing ccSNe are correct, our volumetric rates should be considered lower limits for the ccSN rate in the local universe, as it could be underestimated by at least $\sim 20\%$.
    On the other hand, \citet{2018MNRAS.476.4592D} have shown that SNe II are not subject to large extinctions (also showing that there are no robust methods to correct for host galaxy extinction). If this is the case, our results should reflect the intrinsic volumetric rates of ccSNe, with little or no corrections.
    Finally, our estimates reflect only the visible ccSN rates, as a fraction of massive stars might collapse directly into a black hole, with faint or no optical counterpart, known as failed SNe \citep[e.g.,][]{2008ApJ...684.1336K, 2011ApJ...738..154H, 2013ApJ...769..109L}. For recent limits on the expected fractions of failed SNe, see \citet{2021MNRAS.508..516N} and \citet{2022MNRAS.514.1188B}.
    
    %%%%%%%%%%%%%%%%
    
    \begin{table}
    \renewcommand{\arraystretch}{1.3}
    \small
    \caption{Volumetric rates and fractions of ccSN subtypes. \label{tab:ccSN_rates_sub}}     
    \centering          
    \begin{tabular}{lccc}  
    \hline       
    Type & N$_{\textrm{ccSN}}$ & Fraction $^a$ & Rate \\ 
    &  & [\%] & [$\times 10^{-5}$ yr$^{-1}$ Mpc$^{-3}$ $h^{3}_{70}$]\\
    \hline                    
    II & 104 & $63.2\%^{+14.4\%}_{-13.8\%}$ & $4.3^{+0.8}_{-0.8}$ \\
    SESN & 37 & $32.3\%^{+12.2\%}_{-9.4\%}$ & $2.2^{+0.8}_{-0.6}$\\
    Int. & 31 & $4.4\%^{+1.5\%}_{-1.5\%}$ & $0.3^{+0.1}_{-0.1}$\\
    IIn & 28 & $4.3\%^{+1.9\%}_{-1.6\%}$ & $0.3^{+0.1}_{-0.1}$\\
    Ibn & 3 & $0.45\%^{+0.44\%}_{-0.42\%}$ & $0.031^{+0.031}_{-0.029}$\\
    IIb & 12 & $11.8\%^{+9.9\%}_{-7.1\%}$ & $0.8^{+0.6}_{-0.4}$\\
    Ib & 13 & $11.8\%^{+5.2\%}_{-4.9\%}$ & $0.8^{+0.3}_{-0.3}$\\
    Ic & 8 & $5.8\%^{+4.4\%}_{-3.7\%}$ & $0.4^{+0.3}_{-0.2}$\\
    Ib/c & 2 & $0.47\%^{+0.48\%}_{-0.46\%}$ & $0.033^{+0.033}_{-0.032}$\\
    Ic-BL & 2 & $0.51\%^{+0.55\%}_{-0.32\%}$ & $0.035^{+0.038}_{-0.022}$\\
    \hline                  
    \end{tabular}
    \tablefoot{a. Fraction of total ccSN volumetric rates.}
    \end{table}

    \subsection{The SLSN rate}\label{sec:SLSN}

    \begin{figure}[t!]
    \centerline{
    \includegraphics[scale=0.47]{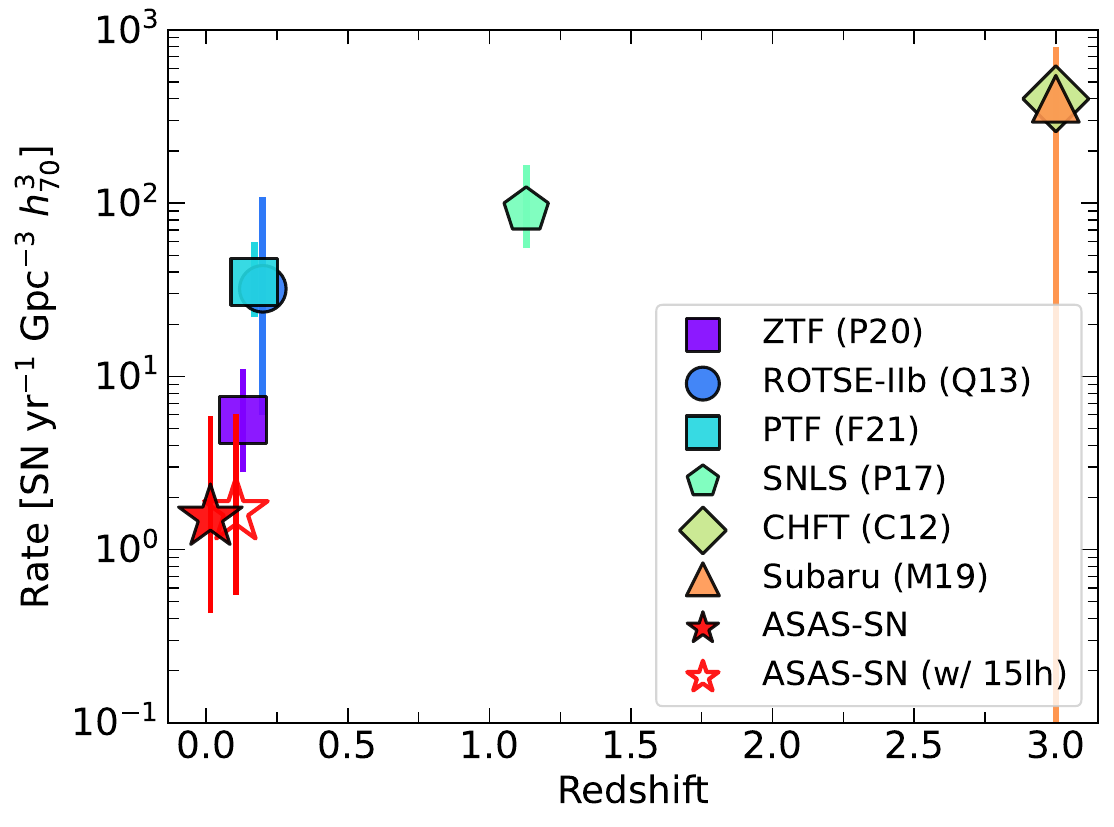}}
    \caption{Volumetric SLSN rates as a function of redshift. Our SLSN rate estimates are shown as the filled ($R_{\textrm{SLSN}}$) and empty ($R_{\textrm{SLSN, w/ 15lh}}$) red stars.
    We also show volumetric SLSN rate estimates from \citetalias{2012Natur.491..228C}, \citetalias{2013MNRAS.431..912Q}, \citetalias{2017MNRAS.464.3568P}, \citetalias{2019ApJS..241...16M}, \citetalias{2020ApJ...904...35P}, and \citetalias{2021MNRAS.500.5142F}.  \label{fig:rate_z_slsn}} 
    \end{figure}

    Between 2014 and 2017, two events detected by the ASAS-SN survey were spectroscopically classified as SLSNe-I: Gaia17biu ($M_V \approx -21.1$ mag, at $z = 0.0307$) and ASASSN-15lh ($M_V \approx -24.7$ mag, at $z = 0.2318$). 
    The latter is, so far, the most luminous optical transient ever detected\footnote{Although more energetic transients have since been discovered \citep{2024arXiv240508855H}.} \citep{2016Sci...351..257D}, and many studies have discussed its true nature \citep{2016NatAs...1E...2L, 2017ApJ...836...25M, 2017MNRAS.466.1428G, 2018ApJ...859..123H, 2018MNRAS.474.3857C, 2018A&A...610A..14K, 2020ApJ...900..121L, 2020MNRAS.497L..13M, 2020MNRAS.498.3730M}. 
    Although the event is spectroscopically similar to SLSNe-I \citep[e.g.,][]{2017MNRAS.466.1428G, 2020MNRAS.498.3730M}, ASASSN-15lh also shows properties consistent with TDEs, as observed in its UV and X-ray emission \citep[e.g., ][]{2017ApJ...836...25M, 2018ApJ...859..123H}, and in its identification as a nuclear event \citep[e.g., ][]{2018A&A...610A..14K}. 
    However, no TDE to date has shown a spectrum similar to ASASSN-15lh.
    Due to this debate, we estimated the volumetric SLSN rate with and without this transient. 

%%%%%%%%%%%

    \begin{table}[t!]
    \renewcommand{\arraystretch}{1.3}
    \small
    \caption{Volumetric SLSN rates. \label{tab:slsn_rates}}     
    \centering          
    \begin{tabular}{cccl}  
    \hline       
    $z$ & N$_{\textrm{SLSN}}$ & Rate & Survey / Ref. \\ 
    &  & [yr$^{-1}$ Gpc$^{-3}$ $h_{70}^3$] & \\
    \hline                    
     $0.001<z<0.1$ & 1 & $1.5^{+4.4}_{-1.1}$ & ASAS-SN $^\dagger$ \\
     $0.001 < z < 0.2318$ & 2 & $1.6^{+4.4}_{-1.1}$ & ASAS-SN (w/15lh) $^\dagger$ \\
     0.11 & 19 & $5.6^{+5.4}_{-2.8}$ & ZTF (\citetalias{2020ApJ...904...35P}) \\
    0.17 & 1 & $32^{+77}_{-26}$ & ROTSE-IIb (\citetalias{2013MNRAS.431..912Q}) \\
    0.17 & 8 & $35^{+25}_{-13}$ & PTF (\citetalias{2021MNRAS.500.5142F}) \\
    1.13 & 3 & $91^{+76}_{-36}$ & SNLS (\citetalias{2017MNRAS.464.3568P}) \\
     $2.0 \leq z \leq 4.0$ & 2 & $\sim 400$ & CHFT (\citetalias{2012Natur.491..228C}) \\
     $2.5 \leq z \leq 3.5$ & 1 & $\sim 400 {\pm400}$& Subaru (\citetalias{2019ApJS..241...16M}) \\
    \hline                  
    \end{tabular}
    \tablefoot{ $^\dagger$ This work.}
    \end{table}
%%%%%%%%%%%

    Considering only Gaia17biu, we find a rate of

    $$ R_{\textrm{SLSN}} = 1.5^{+4.4}_{-1.1} \ \textrm{yr}^{-1} \ \textrm{Gpc}^{-3} \ h^{3}_{70}, $$
    
    \noindent for the redshift interval of $0.001 < z <  0.1 $, which is $0.002\%$ of the total ccSN rate (see Fig. \ref{fig:pie_plot_vol}). 
    By taking only ASASSN-15lh, we estimate a volumetric rate of 15lh-like events of $0.1^{+0.3}_{-0.08} \ \textrm{yr}^{-1} \ \textrm{Gpc}^{-3} \ h^{3}_{70}$. By considering both events, the rate is

    $$ R_{\textrm{SLSN, w/ 15lh}} = 1.6^{+4.4}_{-1.1} \ \textrm{yr}^{-1} \ \textrm{Gpc}^{-3} \ h^{3}_{70}, $$

    \noindent for the redshift range of $0.001 < z <  0.2318$. The uncertainties were estimated using Poisson confidence intervals.
    
    Table \ref{tab:slsn_rates} and Fig. \ref{fig:rate_z_slsn} show both estimated rates together with previous SLSN rate estimates from \citet[][hereafter \citetalias{2012Natur.491..228C}]{2012Natur.491..228C}, \citet[][\citetalias{2013MNRAS.431..912Q}]{2013MNRAS.431..912Q}, \citet[][\citetalias{2017MNRAS.464.3568P}]{2017MNRAS.464.3568P}, \citet[][\citetalias{2019ApJS..241...16M}]{2019ApJS..241...16M}, \citetalias{2020ApJ...904...35P}, and \citetalias{2021MNRAS.500.5142F}.
    Our estimate is lower than those found by \citetalias{2021MNRAS.500.5142F} ($35^{+25}_{-13} \ \textrm{yr}^{-1} \ \textrm{Gpc}^{-3} \ h_{70}^3$, using 8 SLSNe-I) and \citetalias{2020ApJ...904...35P} ($5.6^{+5.4}_{-2.8} \ \textrm{yr}^{-1} \ \textrm{Gpc}^{-3} \ h_{70}^3$, using 19 type I and II SLSNe) for a similar redshift, although it remains consistent with the latter given the uncertainties ($1.2 \ \sigma$).
      These discrepancies may arise from the low numbers in our estimates or differences in sample definitions across different studies.

\subsection{Luminosity functions}

    \begin{figure*}[t]
    \centerline{
    \includegraphics[scale=0.53]{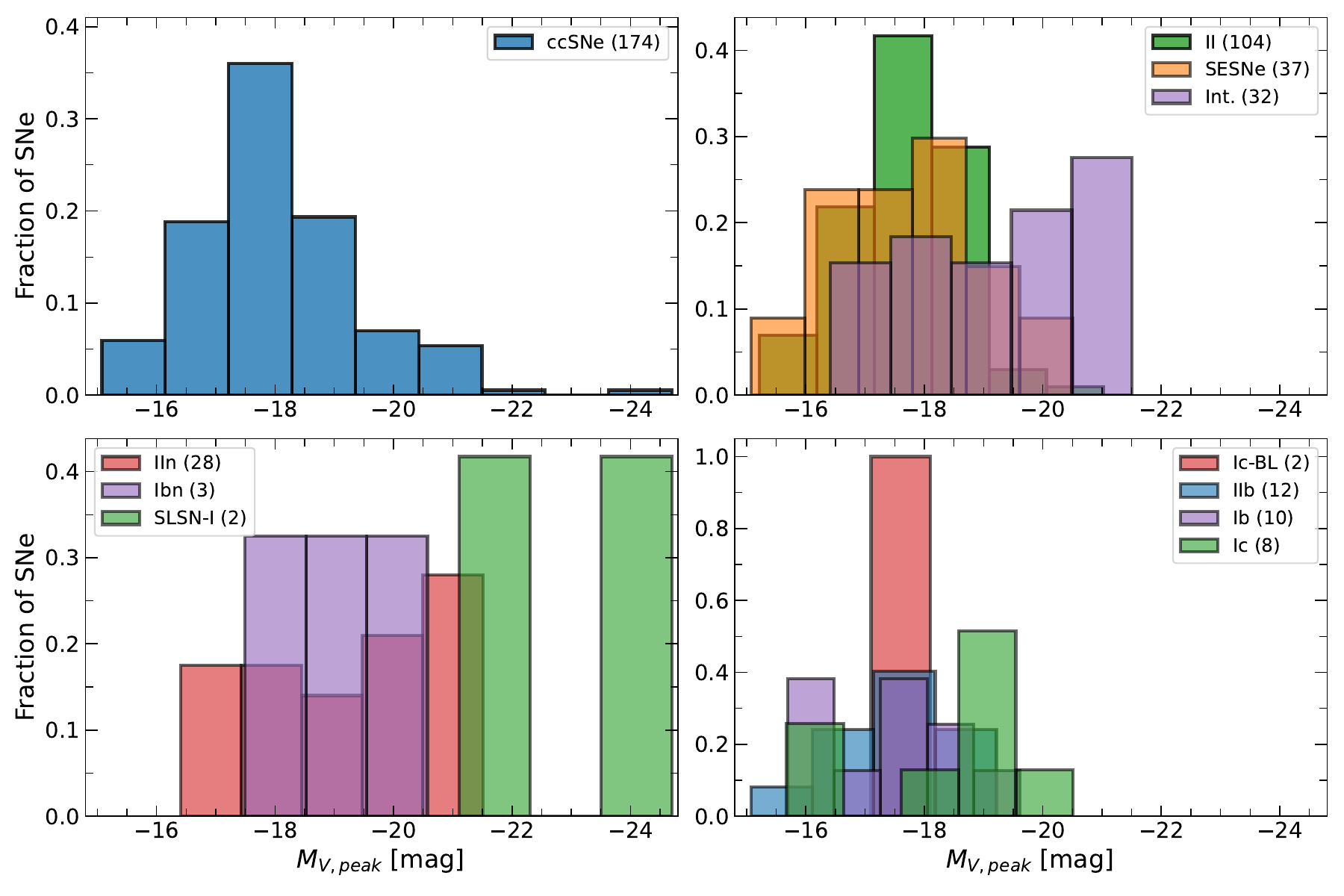}}
    \caption{Distribution of peak absolute magnitudes, $M_{V,peak}$, for the ccSNe in the final sample (top left) and for the different ccSN subtypes (other panels).  \label{fig:abs_mag_hist}} 
    \end{figure*}

    \begin{figure*}[t!]
    \centerline{
    \includegraphics[scale=0.62]{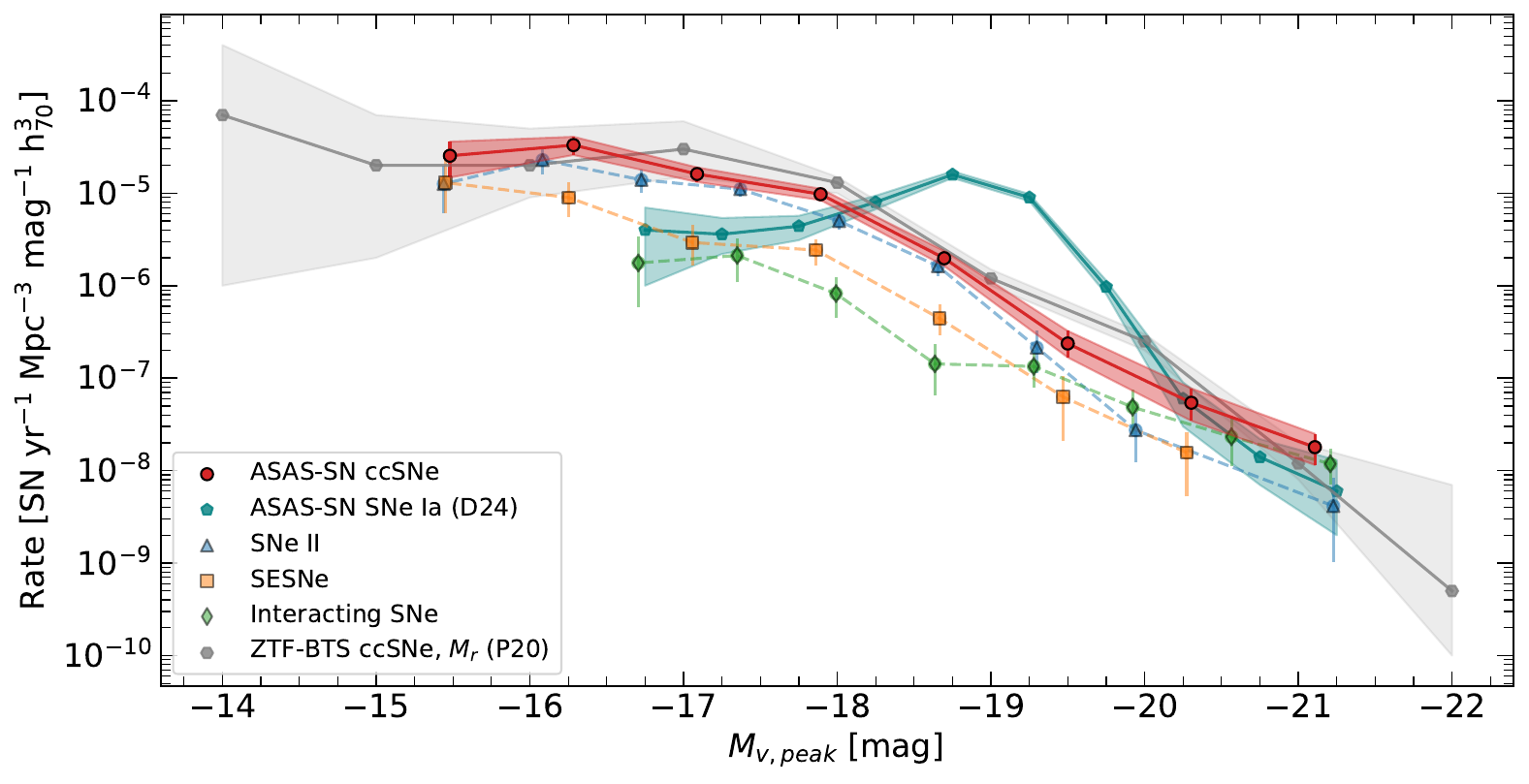}}
    \caption{Luminosity functions of ccSNe. The solid red distribution shows the $V-$band LF considering all the ccSNe in our sample, while the dashed blue, orange, and green lines show the $V-$band LFs for SNe II, SESNe, and Interacting SNe, respectively. We also show the ASAS-SN $V-$band LF for SNe Ia from \citetalias{2023arXiv230611100D}, and the $r-$band ZTF-BTS ccSN LF from \citetalias{2020ApJ...904...35P}. The LFs are corrected for Galactic but not for host galaxy extinction. \label{fig:LF}} 
    \end{figure*}

Figure \ref{fig:abs_mag_hist} shows the distribution of peak $V-$band absolute magnitudes, $M_{V, peak}$, for the full ccSN sample and a range of subsamples. 
The observed values of $M_{V, peak}$ for all ccSNe range from $-15$ to $-22$ mag, with a median of $-17.8$ mag; the unusually bright ASASSN-15lh has $M_{V, peak} \approx -24.7$ mag. SNe II and SESNe have similar distributions, while interacting SNe have slightly more luminous events. Their distributions have medians of $-17.6$, $-17.7$ and $-19.4$ mag, respectively.
Among the 14 non-SLSN-I events with $M_{V, peak} < -20.0$ mag, nine are classified as SNe IIn. Although some studies classify all H-rich SNe with $M_{peak} < -20.0$ mag as SLSNe-II \citep[e.g.,][]{2025A&A...695A.142P}, we simply consider these events as SNe IIn.
ASASSN-14ms is a known luminous SN Ibn with $M_{V, peak} \approx -20.5$ mag \citep{2018MNRAS.475.2344V}, while ASASSN-15nx is a peculiarly luminous SN II with $M_{V, peak} \approx -20.0$ mag \citep[][]{2018ApJ...862..107B}. ASASSN-15um is an unusually bright SESN with $M_{V, peak} \approx -20.5$ mag, possibly indicative of a different explosion mechanism than that of typical SESNe. ASASSN-17rl and -17om (a SN Ib/c and II, respectively) have relatively faint apparent brightness (with $m_{V, peak} \approx 16.6$ and $\approx 16.7$ mag, respectively), which might introduce large uncertainties in the fitting of their light curves and the estimates of their absolute magnitude.

\begin{figure}[t]
    \centerline{
    \includegraphics[scale=0.54]{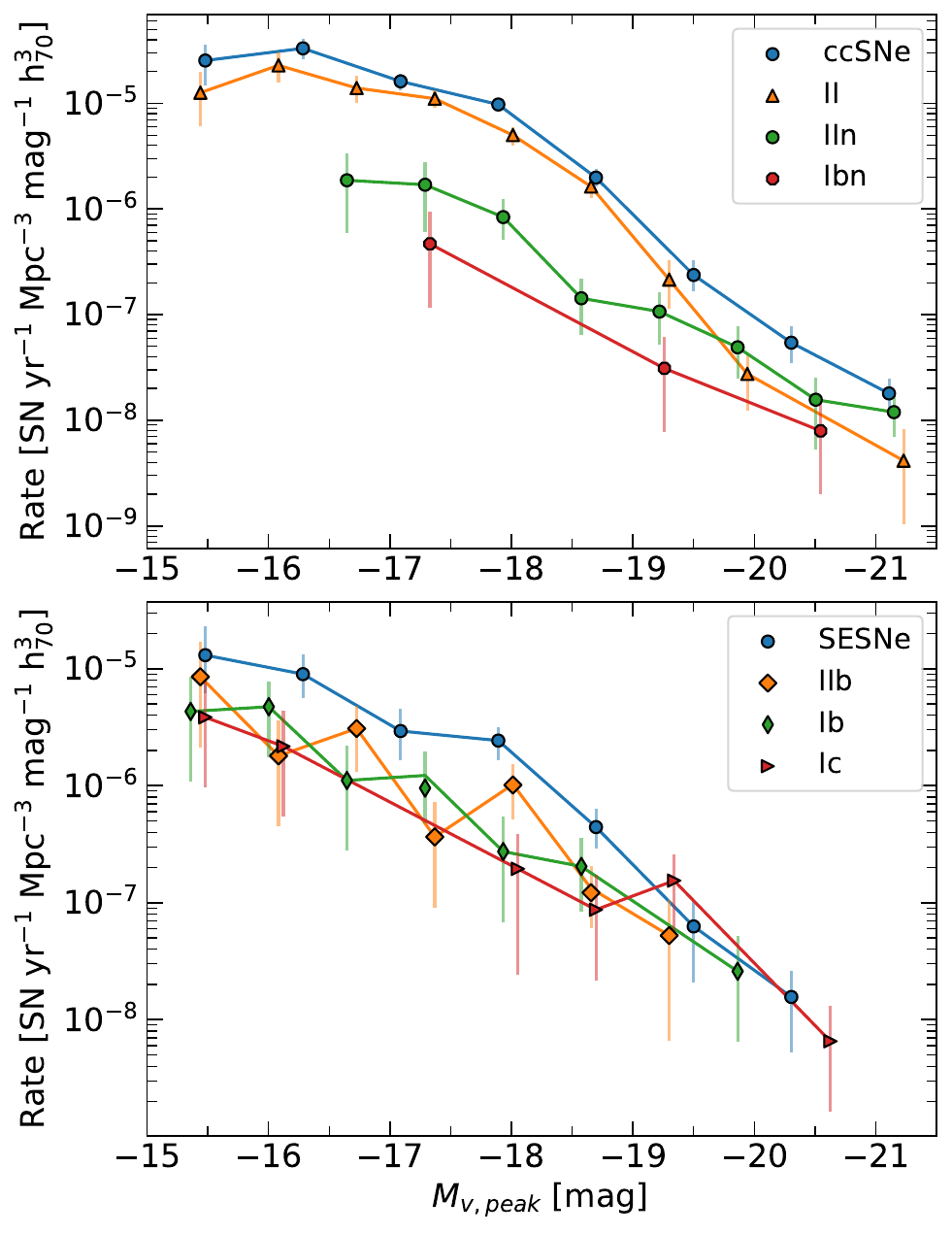}}
    \caption{Luminosity functons for ccSNe and their subtypes. 
    The LFs are corrected only for Galactic extinction. \label{fig:LF_2}} 
    \end{figure}

We estimated the LFs by splitting the ccSN sample into $M_{V, \ peak}$ bins with a width of $0.8$ mag over the range $-15 < M_{V, \ peak} < -21.5$ mag, and calculating the volumetric rate per magnitude using Eq. \ref{eq:rate}.
The resulting ccSN $V-$band LF is shown in Fig. \ref{fig:LF}, together with the $V-$band LFs for SNe II, SESNe, and interacting SNe. Uncertainties are estimated from the 1 $\sigma$ results of $10^3$ bootstrap runs in each absolute magnitude bin, employing a Poisson treatment for small-number statistics. The LF values are reported in Table \ref{tab:LFs}.
The ccSN LF steadily drops with increasing luminosity, with no evidence of a turnover at the faintest luminosities.
SNe II, SESNe, and interacting SNe follow a similar trend. 
Figure \ref{fig:LF_2} again shows the $V-$band LFs of SNe IIn, Ibn, IIb, Ib, and Ic, where a similar trend can be seen.

In Fig. \ref{fig:LF} we also compare our results with the $r-$band ZTF-BTS ccSN LF from \citetalias{2020ApJ...904...35P}. 
Our LF is consistent and follows a very similar trend to the ZTF-BTS LF, although it does not extend to their lower luminosities. Our LF, however, shows smaller uncertainties and is better constrained at the magnitude limits.
Figure \ref{fig:LF} also shows the LF of SNe Ia from \citetalias{2023arXiv230611100D}. 
The ccSNe occur with larger volumetric rates at $M_{V, \ peak} < -18.0$ mag, while SNe Ia are more common at $M_{V, \ peak} \approx -19.0$ mag. The ccSNe and SNe Ia rates are similar for $M_{V, \ peak} > -20.0$ mag.

 As described in Sect. \ref{subsec:ccSN_rates}, we did not apply any host galaxy extinction correction (although Galactic extinction was accounted for). \citetalias{2023arXiv230611100D} show that host extinction should lead to a shift in the estimated LFs in the form $10^{0.6 A_V}$. They compared their LF to \citetalias{2020ApJ...904...35P} LFs in the $r$ band and found a mean host extinction of $E(V-r) \approx 0.2$ mag for SNe Ia. Assuming that the mean host extinction of SNe Ia and ccSNe is similar \cite[e.g.,][]{2025arXiv250307233G}, it should not significantly affect the LFs reported here. 
    
        %%%%%%%%%%%%%%%%

\subsection{The ccSN rate as a function of host galaxy stellar mass} \label{sec:galaxy}

    \begin{figure}[t!]
    \centerline{
    \includegraphics[scale=0.48]{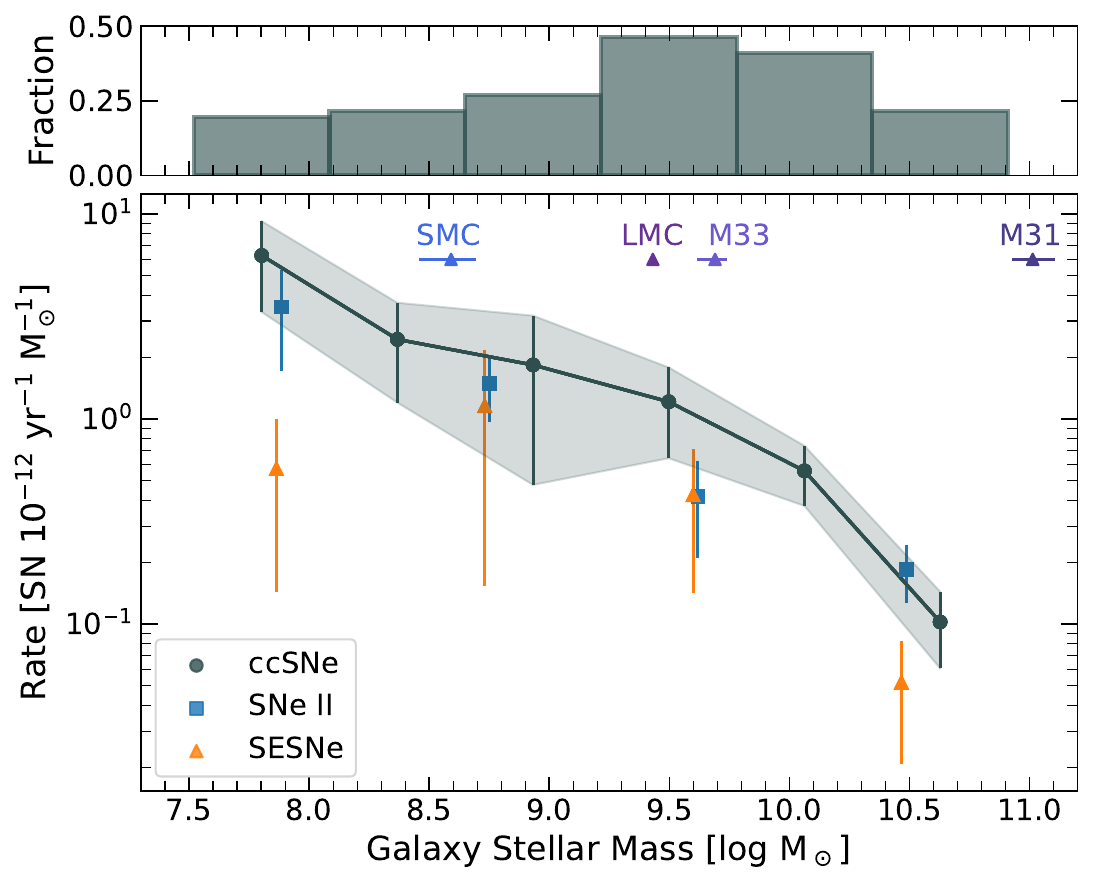}}
    \caption{Specific ccSN rate per unit stellar mass as a function of host galaxy stellar mass.  The blue and orange markers indicate the rates for SNe II and SESNe, respectively. 
    The estimated stellar masses of the Small Magellanic Cloud \citep{2024A&A...681A..15M}, Large Magellanic Cloud \citep{2002AJ....124.2639V}, M33 \citep{2014A&A...572A..23C}, and M31 \citep{2015IAUS..311...82S} are shown at the top of the panel. The top panel shows the distribution of host galaxy stellar mass. \label{fig:host_mass}} 
    \end{figure}

{Finally, we estimated the ccSN rate per unit stellar mass as a function of host galaxy stellar mass. \citet[][]{2021MNRAS.503.3931T} estimated host galaxy masses for 99 of our 174 SNe, limiting the estimates to the north where better photometric data is available. We calculated the rates $R_i$ for this volume and the SN sample in bins of host mass using the same procedures as used to estimate the LFs. We can estimate the stellar mass per unit volume for each mass bin, given a galaxy stellar mass function (GSMF) $dn/dM_*=\phi(M_*)$ as the integral}

\begin{equation}\label{eq:rate_mass}
 \rho_i = \int dM_* M_* \phi(M_*)
\end{equation}

\noindent {over the mass bin, where $M_*$ is the stellar mass of the galaxy. We used the GSMF from \citet{2012MNRAS.421..621B}. 
Figure \ref{fig:host_mass} shows the resulting rate per unit host stellar mass $R_i/\rho_i$ as a function of host mass, together with specific rates for SNe II and SESNe, for masses in the interval $10^{7.5} - 10^{11}$ M$_\odot$.}
    We see a decreasing trend in the ccSN rate per unit stellar mass as stellar mass increases, with higher rates at lower stellar masses. The same is true for both SNe II and SESNe. 
This result highlights the importance of dwarf (log $M_* < 9.0$ M$_\odot$) and lower-mass galaxies in producing ccSNe, as also noted by \citet[][although they analyze the observed distribution of host galaxy stellar mass, and do not correct by the survey completeness]{2021MNRAS.503.3931T}. The Small Magellanic Cloud (SMC)-like galaxies \citep[$M_* = 3.9\pm1.0 \times 10^8$ M$_{\odot}$,][]{2024A&A...681A..15M} produce approximately twice more ccSNe per unit stellar mass than M33-like \citep[$M_* = 4.9^{+0.9}_{-0.7} \times 10^9$ M$_{\odot}$,][]{2014A&A...572A..23C}, and approximately $60$ times more ccSNe per unit stellar mass than M31-like galaxies \citep[$M_* = 1.03^{+0.2}_{-0.1} \times 10^{11}$ M$_{\odot}$,][]{2015IAUS..311...82S}.
    
The trend in Fig. \ref{fig:host_mass} is similar to that observed by \citet{2017ApJ...837..120G}, who estimated the SN rate per unit mass for SNe Ia, SNe II, and SESNe. They also find that the SN rates (for all subtypes) decrease with increasing host galaxy mass and point out that this effect should be generated by the higher SFR present in lower-mass galaxies.
However, \citet{2023ApJ...955L..29P} show that the occurrence of ccSNe increases as metallicity decreases, while being independent of SFR. 
Due to the mass-metallicity relation \citep{2004ApJ...613..898T}, a metallicity effect could explain the behavior seen in Fig. \ref{fig:host_mass}, supporting the findings of \citet{2023ApJ...955L..29P}. 
A similar analysis was done by \citet{2019MNRAS.484.3785B} for SNe Ia from ASAS-SN, who show that the specific rates of SNe Ia increase for lower-mass galaxies. \citet{2022MNRAS.516.1941G} and \citet{2023MNRAS.526.5911J} interpret this result as an effect of higher metallicities and binary fractions in lower-mass galaxies. 

\begin{table*}
\renewcommand{\arraystretch}{1.3}
\small
\caption{The $V-$ band LFs for ccSNe and subtypes. \label{tab:LFs}}     
\centering          
\begin{tabular}{cccccccccc}  
    \hline       
    $M_{V, peak}$ & $R_{\textrm{ccSN}}$ & $R_{\textrm{SESN}}$ & $R_{\textrm{II}}$ & $R_{\textrm{Int.}}$ & $R_{\textrm{IIn}}$ & $R_{\textrm{Ibn}}$ & $R_{\textrm{IIb}}$ & $R_{\textrm{Ib}}$ & $R_{\textrm{Ic}}$\\
    {[mag]} & \multicolumn{9}{c}{{[$10^{-6} \ \textrm{yr}^{-1} \ \textrm{Mpc}^{-3} \ \textrm{mag}^{-1} \ h^{3}_{70} $]}} \\
    \hline                    
{[-20.7, -21.5]} & $0.018^{+0.007}_{-0.006}$ & $0.00$ & $0.003^{+0.003}_{-0.002}$ & $0.013^{+0.006}_{-0.005}$ & $0.013^{+0.006}_{-0.005}$ & $0.00$ & $0.00$ & $0.00$ & $0.00$ \\
{[-19.9, -20.7]} & $0.05^{+0.02}_{-0.02}$ & $0.01^{+0.01}_{-0.01}$ & $0.010^{+0.010}_{-0.007}$ & $0.03^{+0.01}_{-0.01}$ & $0.02^{+0.01}_{-0.01}$ & $0.006^{+0.006}_{-0.005}$ & $0.00$ & $0.00$ & $0.005^{+0.005}_{-0.004}$ \\
{[-19.1, -19.9]} & $0.24^{+0.08}_{-0.07}$ & $0.06^{+0.04}_{-0.04}$ & $0.03^{+0.04}_{-0.02}$ & $0.13^{+0.05}_{-0.05}$ & $0.10^{+0.05}_{-0.04}$ & $0.02^{+0.02}_{-0.02}$ & $0.04^{+0.04}_{-0.03}$ & $0.02^{+0.02}_{-0.01}$ & $0.00$ \\
{[-18.3, -19.1]} & $2.0^{+0.3}_{-0.3}$ & $0.4^{+0.1}_{-0.1}$ & $1.4^{+0.3}_{-0.3}$ & $0.11^{+0.07}_{-0.06}$ & $0.11^{+0.06}_{-0.06}$ & $0.00$ & $0.10^{+0.06}_{-0.05}$ & $0.16^{+0.09}_{-0.09}$ & $0.18^{+0.11}_{-0.10}$ \\
{[-17.5, -18.3]} & $9.7^{+1.5}_{-1.4}$ & $2.3^{+0.7}_{-0.6}$ & $6.3^{+1.1}_{-1.1}$ & $1.0^{+0.5}_{-0.4}$ & $0.6^{+0.3}_{-0.3}$ & $0.3^{+0.3}_{-0.2}$ & $1.1^{+0.5}_{-0.4}$ & $0.21^{+0.21}_{-0.16}$ & $0.15^{+0.15}_{-0.13}$ \\
{[-16.7, -17.5]} & $16.1^{+3.0}_{-2.8}$ & $3.0^{+1.6}_{-1.3}$ & $11.2^{+2.2}_{-2.3}$ & $1.7^{+0.9}_{-0.8}$ & $1.8^{+0.9}_{-0.8}$ & $0.00$ & $0.00$ & $1.7^{+1.2}_{-1.1}$ & $0.00$ \\
{[-15.9, -16.7]} & $33.3^{+7.4}_{-6.8}$ & $9.3^{+4.0}_{-3.7}$ & $22.3^{+6.5}_{-5.6}$ & $1.1^{+1.1}_{-0.8}$ & $1.1^{+1.1}_{-0.8}$ & $0.00$ & $3.8^{+2.1}_{-2.4}$ & $3.7^{+2.3}_{-2.3}$ & $1.7^{+1.7}_{-1.3}$ \\
{[-15.1, -15.9]} & $25.5^{+11.5}_{-9.1}$ & $12.6^{+7.5}_{-6.5}$ & $12.7^{+7.8}_{-7.4}$ & $0.00$ & $0.00$ & $0.00$ & $6.8^{+6.8}_{-5.1}$ & $3.4^{+3.4}_{-2.5}$ & $3.0^{+3.0}_{-2.3}$ \\
    \hline                  
    \end{tabular}
%\tablefoot{}
\end{table*}

\section{Summary and conclusions} \label{sec:conc}

We presented new ccSN rates and LFs, using the ccSNe discovered or recovered by the ASAS-SN survey between 2014 and 2017. 
The spectroscopic classification is nearly complete, as $97\%$ of all transients detected by ASAS-SN over this period were spectroscopically classified. Our final sample is composed of 173 ccSNe in the redshift range $0.001 < z < 0.1$ (plus ASASSN-15lh, an unusually bright transient at $z=0.2318$), with a Galactic latitude $|b| \geq 15^\circ$, a $V-$band absolute magnitude $M_V \leq -14$ mag, and apparent magnitude $m_V < 17.0$ mag (or $m_V < 17.5$ mag, for the ccSNe with long-duration plateaus). 

We used injection recovery simulations to estimate the survey completeness as a function of apparent magnitude for the different ccSN subtypes and light curve shapes. We used these completeness functions to correct for the detectability of each SN and obtain the volumetric rate.
We find a total volumetric rate for ccSNe of $7.0^{+1.0}_{-0.9} \times 10^{-5} \ \textrm{yr}^{-1} \ \textrm{Mpc}^{-3} \ h^{3}_{70}$, at a median redshift of 0.0149. This result is in agreement with the local volumetric rates found in previous studies. 
We obtained rates for the different ccSNe subtypes in our sample, including SNe II, IIn, IIb, Ib, Ic, Ibn, and Ic-BL. 
We find that SNe II comprise $63.2\%$ of all ccSNe, which is $\sim2.8$ times higher than the occurrence rate of SESNe (a volumetric rate of $2.2^{+0.8}_{-0.6} \times 10^{-5} \ \textrm{yr}^{-1} \ \textrm{Mpc}^{-3} \ h^{3}_{70}$), and $\sim22.9$ times higher than the occurrence rate of SNe IIn (a volumetric rate of $3.0^{+1.2}_{-1.0} \times 10^{-6} \ \textrm{yr}^{-1} \ \textrm{Mpc}^{-3} \ h^{3}_{70}$).
We also estimated the volumetric rates for SLSNe to be $1.5^{+4.4}_{-1.1} \ \textrm{yr}^{-1} \ \textrm{Gpc}^{-3} \ h^{3}_{70}$, over the redshift range $0.001 < z < 0.1$, using only the SLSN-I Gaia17biu, and of $1.6^{+4.4}_{-1.1} \ \textrm{yr}^{-1} \ \textrm{Gpc}^{-3} \ h^{3}_{70}$, over the redshift range $0.001 < z < 0.2318$, including the ambiguous superluminous transient ASASSN-15lh. 

We built the LFs of ccSNe and several subtypes in bins of peak $V-$band absolute magnitudes. 
The ccSN rates steadily decline for high luminosities, with no turnover at the faintest luminosities. Type Ia SN rates are higher near $M_{V,peak} \approx -19.0$ mag.
Finally, we estimated the ccSN rate per stellar mass as a function of host galaxy stellar mass. 
The specific rates steadily decline with increasing stellar mass, consistent with previous estimates. Our results show the importance of low-mass galaxies in producing ccSN, with a significantly higher rate per stellar mass than higher mass galaxies, which might be explained by the metallicity dependence on the occurrence of ccSNe \citep[see][]{2023ApJ...955L..29P}.

This paper is part of a series that aim to carefully characterize the LFs and volumetric rates of transients discovered by the ASAS-SN survey, including SNe Ia, ccSNe, as well as such other transients as TDEs. We also plan on using the new ASAS-SN data obtained in the $g$band, for transients discovered after 2018 \citep[the 2018-2020 catalog is presented in][]{2023MNRAS.520.4356N}. Because the $g-$band detections go deeper than $V-$band detections, with a limiting magnitude of $m_g \approx 18.0$ mag, it leads to larger SN samples. The next papers will also focus on a better spectroscopic characterization of SNe and the analysis of other transients discovered by the ASAS-SN survey.

% \section*{Data availability}
% Table \ref{tab:sn_prop} is only available in electronic form at the CDS via anonymous ftp to \url{cdsarc.u-strasbg.fr} (\url{130.79.128.5}) or via \url{http://cdsweb.u-strasbg.fr/cgi-bin/qcat?J/A+A/}.

\begin{acknowledgements}
TP acknowledges the support by ANID through the Beca Doctorado Nacional 202221222222.
The Shappee group at the University of Hawai’i was supported with funds from NSF (grants AST-1908952, AST-1911074, \& AST-1920392) and NASA (grants HST-GO-17087, 80NSSC24K0521, 80NSSC24K0490, 80NSSC24K0508, 80NSSC23K0058, \& 80NSSC23K1431).
CSK and KZS are supported by National Science Foundation grants AST-2307385 and 2407206.
The work of JFB was supported by National Science Foundation Grant No. PHY-2310018.
This work was funded by ANID, Millennium Science Initiative, ICN12\_009. 
This work is supported by the National Natural Science Foundation of China (Grant No. 12133005). S.D. acknowledges the New Cornerstone Science Foundation through the XPLORER PRIZE.
We thank Las Cumbres Observatory and its staff for their continued support of ASAS-SN. ASAS-SN is funded in part by the Gordon and Betty Moore Foundation through grants GBMF5490 and GBMF10501 to the Ohio State University, and also funded in part by the Alfred P. Sloan Foundation grant G-2021-14192. Development of ASAS-SN has been supported by NSF grant AST-0908816, the Mt. Cuba Astronomical Foundation, the Center for Cosmology and AstroParticle Physics at the Ohio State University, the Chinese Academy of Sciences South America Center for Astronomy (CAS- SACA), and the Villum Foundation.
\end{acknowledgements}
   
%--------------------------------------------------------------------
% \bibliographystyle{aa}
\bibliography{ref}

\appendix

\end{document}